\documentclass
[11pt,preprintnumbers,titlepage,nofootinbib,showkeys,showpacs]{revtex4-2}%
\usepackage[latin9]{inputenc}
\usepackage{amsmath}
\usepackage{amssymb}
\usepackage{graphicx}
\usepackage[unicode=true,  bookmarks=false,  breaklinks=false,pdfborder={0 0
1},backref=section,colorlinks=false]{hyperref}
\usepackage{wasysym}
\usepackage{amsfonts}
\usepackage{subfigure}
\usepackage{cleveref}
\usepackage{listings}
\usepackage{xcolor}
\usepackage{array}
\usepackage{url}%
\hypersetup{
colorlinks,linkcolor=blue,citecolor=blue,urlcolor=blue}
\makeatletter
\@ifundefined{textcolor}{}{
\definecolor{BLACK}{gray}{0}
\definecolor{WHITE}{gray}{1}
\definecolor{RED}{rgb}{1,0,0}
\definecolor{GREEN}{rgb}{0,1,0}
\definecolor{BLUE}{rgb}{0,0,1}
\definecolor{CYAN}{cmyk}{1,0,0,0}
\definecolor{MAGENTA}{cmyk}{0,1,0,0}
\definecolor{YELLOW}{cmyk}{0,0,1,0}
}

\makeatother
\begin{document}
\preprint{CTP-SCU/2023001}
\title{Superradiance Instabilities of Charged Black Holes in Einstein-Maxwell-scalar Theory}
\author{Guangzhou Guo$^{a}$}
\email{gzguo@stu.scu.edu.cn}
\author{Peng Wang$^{a}$}
\email{pengw@scu.edu.cn}
\author{Houwen Wu$^{a,b}$}
\email{hw598@damtp.cam.ac.uk}
\author{Haitang Yang$^{a}$}
\email{hyanga@scu.edu.cn}
\affiliation{$^{a}$Center for Theoretical Physics, College of Physics, Sichuan University,
Chengdu, 610064, China}
\affiliation{$^{b}$Department of Applied Mathematics and Theoretical Physics, University of
Cambridge, Wilberforce Road, Cambridge, CB3 0WA, UK}

\begin{abstract}
We study time evolutions of charged scalar perturbations on the background of
a charged hairy black hole, in which the perturbations can be governed by a
double-peak effective potential. By extracting quasinormal modes from the
waveform of scalar perturbations, we discover that some quasinormal modes,
which are trapped in a potential well between two potential peaks, can be
superradiantly amplified. These superradiant modes make the hairy black hole
unstable against charged scalar perturbations. Moreover, it is found that the
superradiant modes arise from the competition between the superradiant
amplification caused by tunneling through the outer potential barrier and the
leakage of modes through the inner potential barrier into the black hole.

\end{abstract}
\maketitle
\tableofcontents

\section{Introduction}

With the advent of successful detections of gravitational waves by LIGO and
Virgo, black hole spectral analysis makes us accessible to test general
relativity in the strong field regime \cite{Abbott:2016blz}. In particular,
the ringdown stage of binary black hole mergers can be precisely modeled by a
superposition of quasinormal modes of remnant black holes \cite{Berti:2007dg}.
Therefore, extracting quasinormal modes from gravitational waves provides a
promising tool to disclose the parameters of black holes, e.g., the black hole
mass, spin and charge \cite{Price:2017cjr,Giesler:2019uxc}. Moreover,
quasinormal modes have been shown to encode geometric information of black
holes
\cite{Ferrari:1984zz,Nollert:1999ji,Yang:2012he,Konoplya:2017wot,Jusufi:2019ltj,Cuadros-Melgar:2020kqn,Qian:2021aju}%
. In the eikonal limit, the real part of quasinormal modes corresponds to the
angular velocity of null circular geodesics, and the imaginary part to the
Lyapunov exponent of the orbits \cite{Cardoso:2008bp}. Additionally,
quasinormal modes play a significant role in determining whether the strong
cosmic censorship is respected for various black hole models
\cite{Cardoso:2017soq,Gan:2019jac,Gan:2019ibg}.

When a black hole spacetime is perturbed, perturbations outside the black hole
are usually damped into the event horizon and spatial infinity, leading to a
discrete set of quasinormal modes with a negative imaginary part
\cite{Berti:2003ud,Berti:2009kk,Konoplya:2011qq,Cook:2016fge,Konoplya:2019hlu}%
. However, the existence of negative energy states can give rise to
superradiance, which allows for energy, charge and angular momentum extraction
from black holes \cite{Brito:2015oca}. Superradiance phenomenon was discovered
when Klein noticed that an beam of electrons can penetrate a step potential
without an exponential suppression \cite{Klein:1929zz}. Intriguingly, a
stationary and axisymmetric spacetime with the presence of an event horizon
has been proven to possess the ergoregion, where negative energy states can
exist \cite{Cardoso:2012zn}. For a rotating black hole, superradiance is the
field-theory version of the Penrose process, in which a test particle can gain
energy from the black hole \cite{Penrose:1969pc}. Analogous to the case of
rotating black holes, superradiance phenomenon was also found for charged
perturbations around a static and charged black hole
\cite{DiMenza:2014vpa,Zhu:2014sya,Benone:2015bst,Corelli:2021ikv}. Moreover, the second law of
black hole thermodynamics implies that waves can extract energy from charged
or rotating black holes if the superradiance condition is fulfilled
\cite{Brito:2015oca,Vicente:2018mxl}.

If extracted energy can pile up outside black holes, superradiance may render
the black hole spacetime unstable against small perturbations. Usually, a
trapping potential well is needed to accumulate superradiant amplifications of
perturbations. On the Kerr background, the effective potential of a massive
scalar field can develop a potential well outside the black hole
\cite{Arvanitaki:2009fg}. The outer potential barrier serves as a mirror to
reflect scalar waves back to the ergoregion to constantly drain energy from
the black hole, resulting in an exponential growth of the scalar field within
the potential well. It is expected that superradiant instabilities may make a
Kerr black hole evolve into a hairy black hole with boson clouds.
Nevertheless, although superradiance can extract up to $29\%$ of the initial
black hole mass \cite{Penrose:1969pc,Hawking:1973uf}, the low energy-density
of boson clouds makes a negligible contribution to the background metric,
which indicates that the formation of boson clouds can be well described at a
linear level \cite{Chia:2022udn}. For a linear perturbation, quasinormal modes
with a positive imaginary part were found, which manifests a superradiance
instability
\cite{Berti:2004ju,Cardoso:2004hs,Cardoso:2004nk,Dias:2022str,Li:2022hkq,Yang:2022uze}%
.

In Reissner-Nordstr\"{o}m (RN) black holes, superradiance instabilities are
absent for a massive charged scalar field since conditions for superradiance
and the existence of a trapping potential well cannot be simultaneously met
\cite{Hod:2012wmy}. To trigger superradiance instabilities, a mirror-like
boundary condition is often imposed for a charged perturbation in the far
field to mimic the outer potential barrier of a massive scalar in the Kerr
background
\cite{Degollado:2013bha,Herdeiro:2013pia,Dolan:2015dha,Sanchis-Gual:2016tcm}.
Notwithstanding, it was found that a potential well of charged scalar
perturbations can be formed outside asymptotically-flat charged black holes in
scalar-tensor Horndeski theory \cite{Kolyvaris:2018zxl}. Nonetheless, the
authors\ made no attempt to search for superradiant modes. Moreover, unstable
superradiant modes were obtained in a charged regular black hole in
\cite{Huang:2015cha}. However, the origin of these modes, i.e., the existence
of a trapping potential well, was not fully discussed.

Recently, a novel class of hairy black holes have been constructed in
Einstein-Maxwell-scalar theory to understand the formation of hairy black
holes \cite{Herdeiro:2018wub}. In particular, the scalar field non-minimally
coupled to the electromagnetic field can trigger the onset of tachyonic
instabilities near the event horizon of RN black holes, inducing scalarized RN
black holes with secondary scalar hair. Further studies on the
Einstein-Maxwell-scalar model have been successively reported in the
literature, e.g., different non-minimal coupling functions
\cite{Fernandes:2019rez,Fernandes:2019kmh,Blazquez-Salcedo:2020nhs}, massive
and self-interacting scalar fields \cite{Zou:2019bpt,Fernandes:2020gay},
horizonless reflecting stars \cite{Peng:2019cmm}, stability analysis of hairy
black holes
\cite{Myung:2018vug,Myung:2019oua,Zou:2020zxq,Myung:2020etf,Mai:2020sac},
higher dimensional scalar-tensor models \cite{Astefanesei:2020qxk},
quasinormal modes of hairy black holes
\cite{Myung:2018jvi,Blazquez-Salcedo:2020jee}, two U(1) fields
\cite{Myung:2020dqt}, quasitopological electromagnetism \cite{Myung:2020ctt},
topology and spacetime structure influences \cite{Guo:2020zqm}, and hairy
black hole solutions in the dS/AdS spacetime
\cite{Brihaye:2019dck,Brihaye:2019gla,Zhang:2021etr,Guo:2021zed}.

Remarkably, we found that scalarized RN black holes can possess two photon
spheres outside the event horizon in certain parameter regions
\cite{Gan:2021pwu}. Subsequently, optical appearances of accretion disks,
luminous celestial spheres and infalling stars in the scalarized RN black hole
background were investigated, showing that an extra photon sphere can lead to
bright rings of different radii and noticeably increasing the flux of observed
accretion disk images \cite{Gan:2021pwu,Gan:2021xdl}, triple higher-order
images of a luminous celestial sphere \cite{Guo:2022muy}, and produce one more
cascade of flashes of an infalling star \cite{Chen:2022qrw}. It is noteworthy
that the existence of two photon spheres outside the event horizon of
asymptotically-flat black holes has also been reported for dyonic black holes
with a quasitopological electromagnetic term and black holes in massive
gravity \cite{Liu:2019rib,deRham:2010kj,Dong:2020odp}. For more details of
black holes with multiple photon spheres, one can refer to \cite{Guo:2022ghl}.

More interestingly, the effective potential of a test neutral scalar field
around the hairy black holes with double photon spheres has been shown to
possess a double-peak profile, which gives rise to long-lived modes trapped in
a potential valley \cite{Guo:2021enm}. In addition, these long-lived modes are
closely related to echo signals observed in black holes with double photon
spheres \cite{Guo:2022umh}. Motivated by the presence of a potential well for
a neutral scalar field, it is natural to explore the existence of a trapping
potential well for a charged scalar field and check superradiance
instabilities against charged scalar perturbations in the hairy black holes.

The rest of the paper is organized as follows. In section \ref{sec:SHBHM},
after scalarized RN black holes are briefly reviewed, we discuss superradiance
for a charged scalar field in the hairy black hole background. Superradiance
instabilities and superradiant quasinormal modes of the hairy black holes are
studied in section \ref{sec:SIHBH}. We finally summarize our results in
section \ref{sec:CONCLUSIONS}. In this paper, we set $16\pi G=1$ hereafter.

\section{Superradiance on Hairy Black Hole Spacetime}

\label{sec:SHBHM}

In this section, we study superradiance of a charged scalar field in static
spherically symmetric charged black holes in an Einstein-Maxwell-scalar model
and show that such fields can extract energy from the black holes.

\subsection{Black Hole Solution}

In the Einstein-Maxwell-scalar model, a scalar filed $\phi$ is minimally
coupled to the metric field and non-minimally coupled to the electromagnetic
field $A_{\mu}$, which is described by the action \cite{Herdeiro:2018wub},
\begin{equation}
S=\int d^{4}x\sqrt{-g}\left[  \mathcal{R}-2\partial_{\mu}\phi\partial^{\mu
}\phi-e^{\alpha\phi^{2}}F_{\mu\nu}F^{\mu\nu}\right]  , \label{eq:Action}%
\end{equation}
where $\mathcal{R}$ is the Ricci scalar, $e^{\alpha\phi^{2}}$ is the coupling
function, and $F_{\mu\nu}=\partial_{\mu}A_{\nu}-\partial_{\nu}A_{\mu}$ is the
electromagnetic field strength tensor. For a static and spherically symmetric
black hole solution,
\begin{align}
ds^{2}  &  =-N(r)e^{-2\delta(r)}dt^{2}+\frac{1}{N(r)}dr^{2}+r^{2}\left(
d\theta^{2}+\sin^{2}\theta d\varphi^{2}\right)  ,\nonumber\\
A_{\mu}dx^{\mu}  &  =\Phi(r)dt\text{ and}\ \phi=\phi(r), \label{eq:HBH}%
\end{align}
one can obtain the corresponding equations of motion,
\begin{align}
N^{\prime}(r)  &  =\frac{1-N(r)}{r}-\frac{Q^{2}}{r^{3}e^{\alpha\phi^{2}(r)}%
}-rN(r)\left[  \phi^{\prime}(r)\right]  ^{2},\nonumber\\
\left[  r^{2}N(r)\phi^{\prime}(r)\right]  ^{\prime}  &  =-\frac{\alpha
Q^{2}\phi(r)}{r^{2}e^{\alpha\phi^{2}(r)}}-r^{3}N(r)\left[  \phi^{\prime
}(r)\right]  ^{3},\nonumber\\
\delta^{\prime}(r)  &  =-r\left[  \phi^{\prime}(r)\right]  ^{2},\label{eq:EOM}%
\\
\Phi^{\prime}(r)  &  =\frac{Q}{r^{2}e^{\alpha\phi^{2}(r)}}e^{-\delta
(r)},\nonumber
\end{align}
where the integration constant $Q$ denotes the black hole electric charge, and
primes represent derivatives with respect to $r$.

\begin{figure}[ptb]
\begin{centering}
\includegraphics[scale=1]{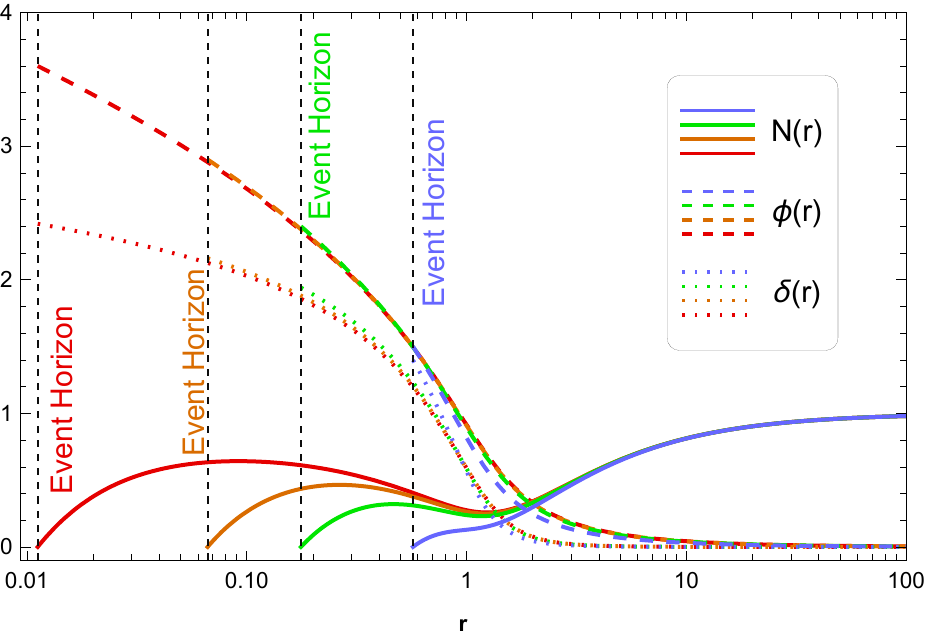}
\par\end{centering}
\caption{Hairy black hole solutions with $\alpha=0.88$ for $Q=1.0365$ (blue
lines), $Q=1.0629$ (green lines), $Q=1.0691$ (orange lines) and $Q=1.0717$
(red lines). The area of the event horizon shrinks as the black hole charge
increases.}%
\label{Fig: Hairy BH}%
\end{figure}

To solve for the black hole solution from eqn. $\left(  \ref{eq:EOM}\right)
$, appropriate boundary conditions should be taken into account. At the event
horizon $r_{h}$, we impose that
\begin{equation}
N(r_{h})=0\text{, }\delta(r_{h})=\delta_{0}\text{, }\phi(r_{h})=\phi
_{0}\text{, }V(r_{h})=\Phi_{0}\text{,} \label{eq:rh condition}%
\end{equation}
where $\Phi_{0}$ is the electrostatic potential, and $\delta_{0}$ and
$\phi_{0}$ are used to characterize a specific black hole solution. In
particular, $\phi_{0}=\delta_{0}=0$ corresponds to a scalar-free black hole
solution with $\phi=0$, i.e., a RN black hole. Nevertheless, when nonzero
values of $\phi_{0}$ and $\delta_{0}$ are adopted, a hairy black hole solution
with a non-trivial scalar field $\phi$ can be obtained. At spatial infinity,
the black hole solution has the asymptotic expansions,
\begin{equation}
N(r)=1-\frac{2M}{r}+\frac{Q^{2}+Q_{s}^{2}}{r^{2}}+...\text{,}\ \delta
(r)=\frac{Q_{s}^{2}}{2r^{2}}+...\text{, }\phi(r)=\frac{Q_{s}}{r}+...\text{,
}\Phi(r)=-\frac{Q}{r}+...\text{,} \label{eq:infinity condition}%
\end{equation}
where $M$ is the black hole mass, and $Q_{s}$ is the scalar charge. In this
paper, we use a shooting method built in the $NDSolve$ function of
$Wolfram\text{ }\circledR Mathematica$ to find hairy black hole solutions
satisfying the boundary conditions $\left(  \ref{eq:rh condition}\right)  $
and $\left(  \ref{eq:infinity condition}\right)  $. Thanks to the scaling
symmetry, we can set $M=1$ without loss of generality. In FIG.
\ref{Fig: Hairy BH}, we present the functions $N(r)$, $\delta(r)$ and
$\phi(r)$ of the black hole solutions with $\alpha=0.88$ for various values of
$Q$.

\subsection{Superradiance of a Charged Scalar Field}

We now study a massless charged scalar field $\Psi$ of charge $q$ perturbing
the hairy black hole spacetime, which is governed by the Klein-Gordon
equation,
\begin{equation}
\left(  \nabla_{\mu}-iqA_{\mu}\right)  \left(  \nabla^{\mu}-iqA^{\mu}\right)
\Psi=0. \label{eq:KG-eq}%
\end{equation}
Since the background spacetime is spherically symmetric, we decompose the
scalar perturbation $\Psi$ in terms of spherical harmonic functions,
\begin{equation}
\Psi=\sum_{lm}\psi_{lm}\left(  t,r\right)  Y_{lm}\left(  \theta,\varphi
\right)  /r.
\end{equation}
The Klein-Gordon equation is therefore reduced to%
\begin{equation}
\frac{\partial^{2}\psi(t,r)}{\partial t^{2}}-\frac{\partial^{2}\psi
(t,r)}{\partial x^{2}}-2iq\Phi(r)\frac{\partial\psi(t,r)}{\partial t}-\left[
q^{2}\Phi^{2}(r)-V(r)\right]  \psi(t,r)=0, \label{eq:scalar-eq}%
\end{equation}
where the indices $l$ and $m$ are suppressed for simplicity, the tortoise
coordinate $x$ is defined as $dx/dr=e^{\delta(r)}/N(r)$, and the scalar
potential is defined as
\begin{equation}
V(r)=\frac{N(r)}{r^{2}e^{2\delta(r)}}\left[  1-N(r)-\frac{Q^{2}}%
{r^{2}e^{\alpha\phi^{2}(r)}}+l\left(  l+1\right)  \right]  .
\end{equation}
Note that $x=-\infty$ and $+\infty$ correspond to $r=r_{h}$ and $\infty$,
respectively. Given an initial scalar perturbation, solving the partial
differential equation $\left(  \ref{eq:scalar-eq}\right)  $ gives the
evolution of the scalar perturbation in the hairy black hole.

\begin{figure}[ptb]
\includegraphics[scale=0.58]{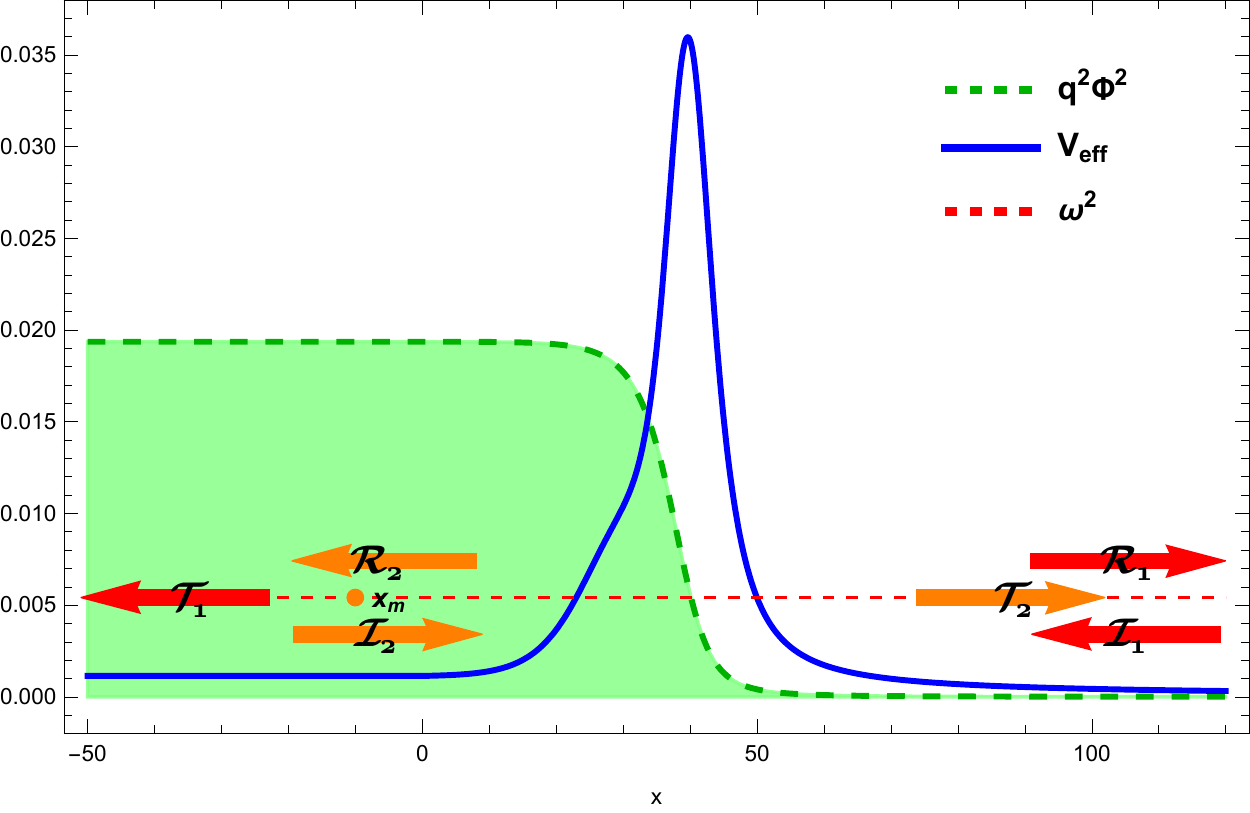}$\quad$%
\includegraphics[scale=0.78]{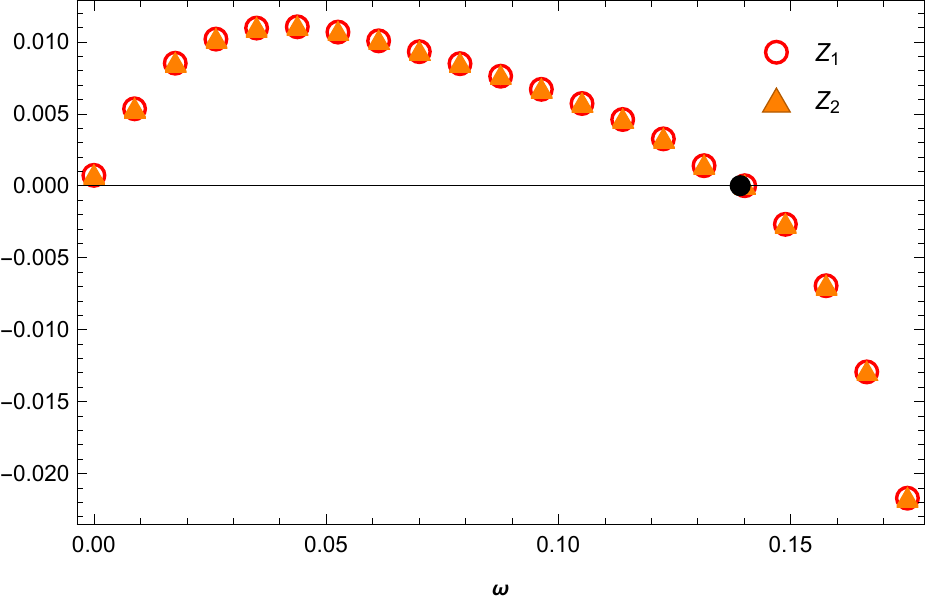}
\par
\includegraphics[scale=0.58]{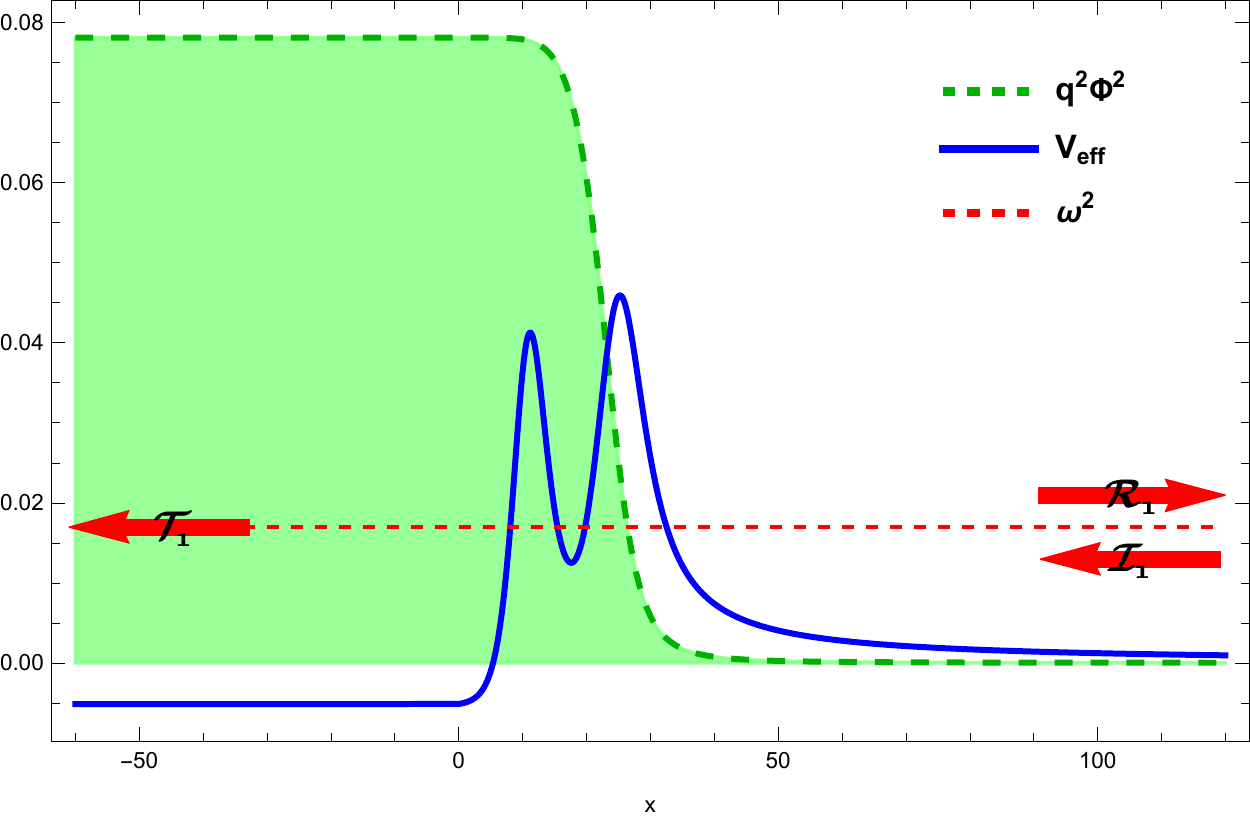}$\quad$%
\includegraphics[scale=0.78]{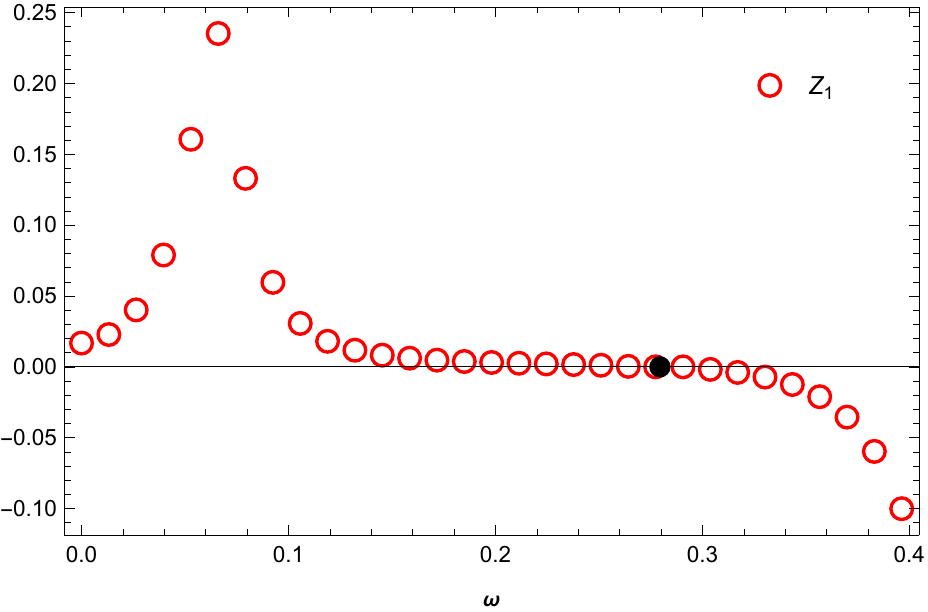}\caption{\textbf{Upper Row}:
Superradiance of a massless scalar field with $l=0$ and charge $q=0.15$ in the
hairy black hole with $\alpha=0.88$ and $Q=1.0365$. The left panel exhibits
the effective potential of the scalar field with $\omega=0.0736$ and the
$q^{2}\Phi^{2}$ line bounding a green region, in which the phase and group
velocities of the scalar wave are in opposite direction. The existence of the
green region can induce superradiance for incident scalar waves travelling in
both directions. Amplification factors $Z_{1}=\frac{\left\vert \mathcal{R}%
_{1}\right\vert ^{2}}{\left\vert \mathcal{I}_{1}\right\vert ^{2}}-1$ and
$Z_{2}=\frac{\left\vert \mathcal{R}_{2}\right\vert ^{2}}{\left\vert
\mathcal{I}_{2}\right\vert ^{2}}-1$ are plotted as a function of the frequency
$\omega$ in the right panel, which shows that $Z_{1}=Z_{2}$. The superradiance
with $Z_{1}=Z_{2}>0$ occurs when $\omega<\omega_{\text{up}}\simeq0.1391$,
which is denoted by a black dot. \textbf{Lower Row}: Superradiance of a
massless scalar field with $l=0$ and charge $q=0.3$ in the hairy black hole
with $\alpha=0.88$ and $Q=1.0629$. The threshold for superradiance with
$Z_{1}>0$ to occur is at $\omega=\omega_{\text{up}}\simeq0.2795$, marked by a
black dot. Due to a larger value of $qQ$, the amplification factor $Z_{1}$
significantly increases. The effective potential presents a double-peak
structure with a potential well, which provides an arena for superradiance
instabilities.}%
\label{Fig:tunnel}%
\end{figure}

Alternatively, one can study superradiance of the scalar field in the
frequency domain. By Fourier transforming $\psi(t,r)=\int d\omega e^{-i\omega
t}\psi(r)/2\pi$, eqn. $\left(  \ref{eq:scalar-eq}\right)  $ can be written in
the Schr\"{o}dinger-like form,%
\begin{equation}
\left[  -\frac{d^{2}}{dx^{2}}+V_{\text{eff}}(\omega,r)\right]  \psi
(r)=\omega^{2}\psi(r), \label{eq:freq-eq}%
\end{equation}
where we define a frequency-dependent effective potential for the scalar
perturbation,
\begin{equation}
V_{\text{eff}}(\omega,r)=V(r)-2q\omega\Phi(r)-q^{2}\Phi^{2}(r).
\end{equation}
It can show that $V_{\text{eff}}(\omega,\infty)=0$, and $V_{\text{eff}}%
(\omega,r_{h})=\omega^{2}-k_{h}^{2}$ with $k_{h}=\omega+q\Phi(r_{h})$. Since
$V_{\text{eff}}(\omega,r)$ is constant at the boundaries, a solution $\psi
_{1}$ to eqn. $\left(  \ref{eq:freq-eq}\right)  $ can have the following
asymptotic behavior,
\begin{align}
\psi_{1}(x)  &  \sim\mathcal{T}_{1}e^{-ik_{h}x}\text{ for }x\rightarrow
-\infty\nonumber\\
\psi_{1}(x)  &  \sim\mathcal{I}_{1}e^{-i\omega x}+\mathcal{R}_{1}e^{+i\omega
x}\text{ for }x\rightarrow+\infty,
\end{align}
which describes an incident wave of amplitude $\mathcal{I}_{1}$ from spatial
infinity scattering off the effective potential to produce a reflected wave of
amplitude $\mathcal{R}_{1}$ and a transmitted wave of amplitude $\mathcal{T}%
_{1}$ (see FIG. \ref{Fig:tunnel}). It should be emphasized that when
$\omega<-q\Phi(r_{h})$, the phase velocity of the transmitted wave is
positive, describing a wave moving away from the horizon.\ Nevertheless, the
group velocity of this wave is negative, e.g., $v_{g}=-d\omega/dk_{h}=-1$,
indicating that any wavepacket consisting of such modes still moves toward the horizon.

As discussed in \cite{Hod:2012wmy,Zhu:2014sya,Huang:2015cha}, the Wronskian
identity can be applied to deriving a relation between the scattering
coefficients. The Wronskian is constructed by two linearly independent
solutions $f_{1}$ and $f_{2}$ to eqn. $\left(  \ref{eq:freq-eq}\right)  $,
\begin{equation}
W=f_{1}\frac{df_{2}}{dx}-f_{2}\frac{df_{1}}{dx},
\end{equation}
which is independent of the spatial coordinate. Therefore, the Wronskian
evaluated at the event horizon equals that evaluated at spatial infinity,%
\begin{equation}
\left\vert \mathcal{R}_{1}\right\vert ^{2}-\left\vert \mathcal{I}%
_{1}\right\vert ^{2}=-\frac{k_{h}}{\omega}\left\vert \mathcal{T}%
_{1}\right\vert ^{2},
\end{equation}
where we set $f_{1}=\psi_{1}$ and $f_{2}=\psi_{1}^{\ast}$. It is evident that,
when $0<\omega<-q\Phi(r_{h})\equiv\omega_{\text{up}}$, the reflected wave is
superradiantly amplified, i.e., $\left\vert \mathcal{R}_{1}\right\vert
>\left\vert \mathcal{I}_{1}\right\vert $. To quantify the superradiance, one
can define an amplification factor%
\begin{equation}
Z_{1}=\frac{\left\vert \mathcal{R}_{1}\right\vert ^{2}}{\left\vert
\mathcal{I}_{1}\right\vert ^{2}}-1. \label{eq:Z1}%
\end{equation}

The amplification factor $Z_{1}$ with a given $\omega$ can be computed by
numerically integrating eqn. $\left(  \ref{eq:freq-eq}\right)  $. The right
panels of FIG. \ref{Fig:tunnel} display the amplification factor $Z_{1}$ as a
function of the frequency $\omega$ for $l=0$ waves with different black hole
charge $Q$ and scalar field charge $q$. The black dots denote the
superradiance threshold, $Z_{1}=0$, which is found to be consistent with
$\omega=\omega_{\text{up}}$. When $\omega<\omega_{\text{up}}$, superradiance
is expected to occur, i.e., $Z_{1}>0$. Moreover, it shows that a larger value
of $qQ$ would substantially increase the amplification factor of the scalar
field. In addition, the corresponding effective potential $V_{\text{eff}%
}(\omega,x)$ and the function $q^{2}\Phi^{2}(x)$ are presented in the left
panels of FIG. \ref{Fig:tunnel}. Interestingly, the lower-left panel shows
that there can exist two peaks in the effective potential for a large enough
black hole charge. As discussed below, the potential well between the two
peaks can temporarily trap scalar perturbations, and hence plays a pivotal
role in superradiant instabilities of black holes.

In hairy black holes with a double-peak effective potential, if the scalar
field is perturbed between the two potential peaks, an outgoing scalar wave
will be incident on the outer potential peak and get scattered. To investigate
such case, we consider that an outgoing beam of scalar scatters off a
potential barrier, which produces reflected and transmitted waves traveling
towards the event horizon and spatial infinity, respectively. As illustrated
in the upper-left panel of FIG. \ref{Fig:tunnel}, this tunneling process is
described by a solution $\psi_{2}$ to eqn. $\left(  \ref{eq:freq-eq}\right)  $
with the asymptotic behavior,
\begin{align}
\psi_{2}(x)  &  \sim\mathcal{I}_{2}e^{ik_{m}x}+\mathcal{R}_{2}e^{-ik_{m}%
x}\text{ near\ }x=x_{m},\nonumber\\
\psi_{2}(x)  &  \sim\mathcal{T}_{2}e^{i\omega x}\text{ for }x\rightarrow
+\infty,
\end{align}
where $x_{m}$ is some reference point, and $k_{m}=-\sqrt{[\omega+q\Phi\left(
x_{m}\right)  ]^{2}-V\left(  x_{m}\right)  }$ and $\sqrt{[\omega+q\Phi\left(
x_{m}\right)  ]^{2}-V\left(  x_{m}\right)  }$ for $\omega<-q\Phi\left(
x_{m}\right)  $ and $\omega>-q\Phi\left(  x_{m}\right)  $, respectively. Since
$d\omega/dk_{m}>0$, the ingoing reflected wave of amplitude $\mathcal{R}_{2}$
and the outgoing incident wave of amplitude $\mathcal{I}_{2}$ have negative
and positive group velocities, respectively. Evaluating the Wronskian identity
for $\psi_{2}$ and $\psi_{2}^{\ast}$ at $x=x_{m}$ and $+\infty$ gives the
relation
\begin{equation}
\left\vert \mathcal{R}_{2}\right\vert ^{2}-\left\vert \mathcal{I}%
_{2}\right\vert ^{2}=-\frac{\omega}{k_{m}}\left\vert \mathcal{T}%
_{2}\right\vert ^{2},
\end{equation}
which shows that the reflected wave has a larger amplitude than that of the
incident wave if $0<\omega<-q\Phi(x_{m})$. Similarly, one can define an
amplification factor $Z_{2},$
\begin{equation}
Z_{2}=\frac{\left\vert \mathcal{R}_{2}\right\vert ^{2}}{\left\vert
\mathcal{I}_{2}\right\vert ^{2}}-1. \label{eq:Z2}%
\end{equation}
If $x_{m}$ is far away from the potential barrier, i.e., $k_{m}\simeq k_{h}$,
evaluating the Wronskian for the independent solutions $\psi_{1}$ and
$\psi_{2}$ yields
\begin{equation}
k_{h}\mathcal{I}_{1}\mathcal{T}_{2}=-\omega\mathcal{I}_{2}\mathcal{T}_{1}.
\label{eq:T1T2}%
\end{equation}
Using eqns. $\left(  \ref{eq:Z1}\right)  $, $\left(  \ref{eq:Z2}\right)  $ and
$\left(  \ref{eq:T1T2}\right)  $, one can easily infer that $Z_{1}=Z_{2}$.

In the upper-right panel of FIG. \ref{Fig:tunnel}, we present the
amplification factor $Z_{2}$ for a charged scalar perturbation with $q=0.15$
and $l=0$ in the hairy black hole with $\alpha=0.88$ and $Q=1.0365$. As
expected, it shows that $Z_{2}>$ $0$ when $0<\omega<\omega_{\text{up}}$, and
$Z_{1}=Z_{2}$. Furthermore, green regions of FIG. \ref{Fig:tunnel} are bounded
by $q^{2}\Phi^{2}\left(  x\right)  $, below which scalar waves have
$0<\omega<-q\Phi\left(  x\right)  $. Our results reveal that superradiance of
scalar waves can occur when transmitted waves enter or leave the green
regions. Indeed for a scalar mode of frequency $\omega$, one has%
\begin{equation}
\omega=-\left(  p_{\mu}+qA_{\mu}\right)  \eta^{\mu}=-p_{\mu}\eta^{\mu}-q\Phi,
\end{equation}
where $p_{\mu}$ is the four momentum, and $\eta^{\mu}=\left(  \partial
/\partial t\right)  ^{\mu}$ is a Killing vector. The local energy of the mode
measured by a static observer with four-velocity $u^{\mu}=N^{-1/2}e^{\delta
}\eta^{\mu}$ is
\begin{equation}
E=-p_{\mu}u^{\mu}=N^{-1/2}e^{\delta}\left(  \omega+q\Phi\right)  ,
\end{equation}
which becomes negative in the green regions. Therefore, if an outgoing scalar
wave tunnels through the potential barrier, the transmitted wave leaves the
green region and can transport positive energy to spatial infinity, which
amplifies negative energy states in the green region. Additionally, when the
transmitted part of an incoming scalar wave propagating from spatial infinity
enters the green region, negative energy would be carried away to the black
hole, and hence the outgoing reflected wave gets superradiantly amplified.

\section{ Superradiant Instabilities of Hairy Black Holes}

\label{sec:SIHBH}

As previously mentioned, a particularly important feature of the hairy black
hole solution in the Einstein-Maxwell-scalar model is that the presence of a
potential well, which are able to trap and amplify a charged scalar field, and
hence trigger a superradiant instability. In this section, we investigate the
superradiant instability of hairy black holes by computing quasinormal modes
of the charged scalar field.

To obtain quasinormal modes, ingoing and outgoing boundary conditions are
imposed at the event horizon and spatial infinity,
\begin{align}
\psi &  \sim e^{-i[\omega+q\Phi(r_{h})]x},\quad\text{for }x\rightarrow
-\infty,\nonumber\\
\psi &  \sim e^{i\omega x},\quad\text{for }x\rightarrow+\infty,
\label{eq:QNMBC}%
\end{align}
respectively, which selects a discrete set of quasinormal modes of frequency
$\omega_{n}$ with the overtone number $n=0,1,2\ldots.$ Owing to the boundary
conditions $\left(  \ref{eq:QNMBC}\right)  $, the quasinormal frequencies
$\omega$ are generically complex, $\omega=\omega^{R}+i\omega^{I}$. Especially,
a quasinormal mode with $\omega_{I}>0$ corresponds to an instability with the
instability time scale $1/\omega^{I}$.

Here, we use time-domain analysis to extract quasinormal frequencies from time
evolutions of the scalar field $\psi(t,x)$, which is governed by eqn. $\left(
\ref{eq:scalar-eq}\right)  $. Specifically, we employ a finite difference
numerical scheme to discretize the partial differential equation $\left(
\ref{eq:scalar-eq}\right)  $ \cite{Zhu:2014sya,Huang:2015cha}. Denoting
$\psi_{i,j}=\psi\left(  i\Delta t,j\Delta x\right)  $, $\Phi_{j}=\Phi\left(
j\Delta x\right)  $ and $V_{j}=V\left(  j\Delta x\right)  $, the evolution of
$\psi\left(  t,x\right)  $ is determined by
\begin{equation}
\psi_{i+1,j}=-\frac{\left(  1+iq\Phi_{j}\Delta t\right)  \psi_{i-1,j}%
}{1-iq\Phi_{j}\Delta t}+\frac{\Delta t^{2}}{\Delta x^{2}}\frac{\psi
_{i,j+1}+\psi_{i,j-1}}{1-iq\Phi_{j}\Delta t}+\left[  2-2\frac{\Delta t^{2}%
}{\Delta x^{2}}-\Delta t^{2}(V_{j}-q^{2}\Phi_{j}^{2})\right]  \frac{\psi
_{i,j}}{1-iq\Phi_{j}\Delta t}, \label{eq:psi evolution}%
\end{equation}
where we set $\Delta t/\Delta x=1/2$ to avoid numerical divergence in the
simulation \cite{Konoplya:2013rxa,Zhu:2014sya}. Additionally, we assume that
the initial condition of $\psi\left(  t,x\right)  $ is a Gaussian wavepacket,
$\psi\left(  t=0,x\right)  =\exp\left[  -\left(  x-a\right)  ^{2}%
/2b^{2}\right]  $ and $\psi\left(  t<0,x\right)  =0$.

After the late-time waveform of $\psi(t,x)$ is obtained, quasinormal modes can
be extracted via the Prony method
\cite{Berti:2007dg,Zhu:2014sya,Huang:2015cha}. In particular, the late-time
waveform can be expressed as the superposition of a set of quasinormal modes,
\begin{equation}
\psi(t,x)=\sum_{i=1}^{p}C_{i}e^{-i\omega_{i}(t-t_{0})}.
\label{eq:late-time waveform}%
\end{equation}
For a given integer $n$, the value of $\psi(t_{0}+n\Delta t)$ can be
numerically obtained from eqn. $\left(  \ref{eq:psi evolution}\right)  $. By
fitting the late-time waveform $\left(  \ref{eq:late-time waveform}\right)  $
with $\psi(t_{0}+n\Delta t)$ in the time interval between $t_{0}$ and
$t_{0}+N\Delta t$, one can perform the Prony method to obtain the coefficients
$C_{i}$ and the quasinormal frequencies $\omega_{i}$.

\subsection{Single-peak Potential}

\begin{figure}[ptb]
\includegraphics[scale=0.6]{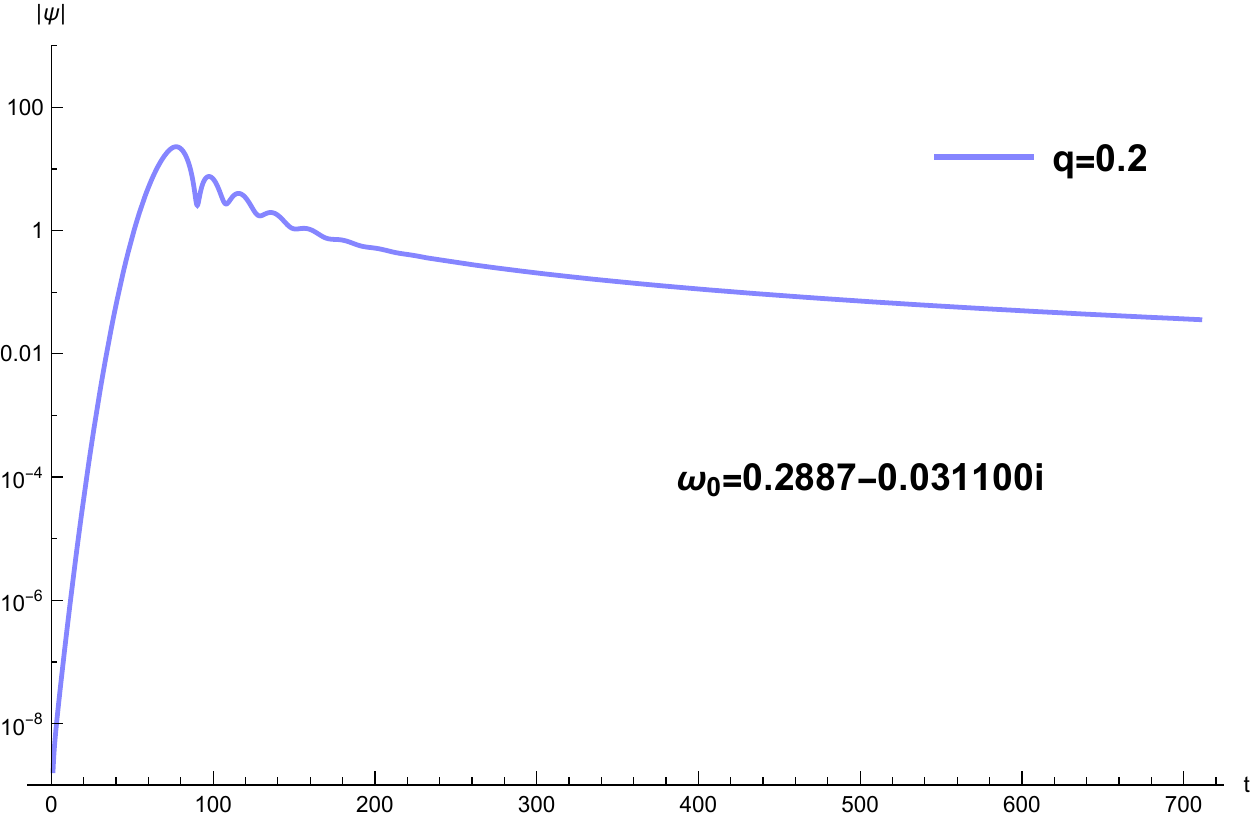}$\quad$%
\includegraphics[scale=0.6]{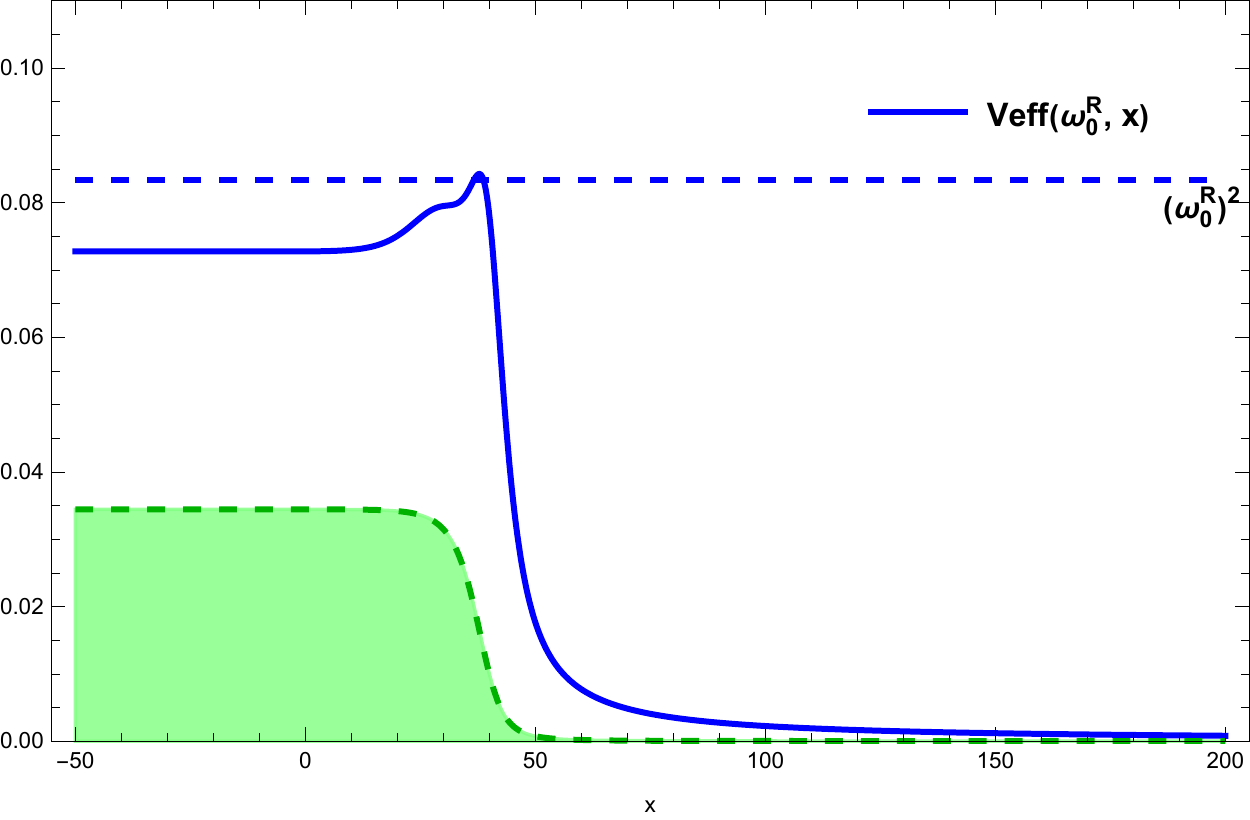}\caption{\textbf{Left Panel}: Time
evolution of the $q=0.2$ and $l=0$ scalar field in the hairy black hole with
$\alpha=0.88$ and $Q=1.0365$ The observer is located at $x=114$. At late
times, the scalar field decays exponentially, showing that the superradiance
instability is absent. The fundamental quasinormal mode $\omega_{0}$ is
extracted from the late-time waveform of the scalar field. \textbf{Right
Panel}: The effective potential $V_{\text{eff}}(\omega_{0}^{R},x)$ of the
fundamental quasinormal mode has only one maximum. }%
\label{Fig:RN type}%
\end{figure}

We first check superradiant instabilities of hairy black holes against the
charged scalar field, whose effective potential has only one maximum. The left
panel of FIG. \ref{Fig:RN type} exhibits the time evolution of the scalar
field with $q=0.2$ and $l=0$ in the hairy black hole with $\alpha=0.88$ and
$Q=1.0365$. The late-time waveform is an exponential decay, and hence the
hairy black hole is stable against the charged scalar perturbation. As
anticipated, the imaginary part of the extracted quasinormal mode is negative.
The effective potential of the quasinormal mode is presented in the right
panel of FIG. \ref{Fig:RN type}. Similar to RN black holes, the absence of
superradiant instabilities is attributed to the non-existence of a trapping
potential well outside the black hole \cite{Hod:2012wmy}.

\subsection{Potential with Two Well-separated Peaks}

\begin{figure}[ptb]
\includegraphics[scale=0.6]{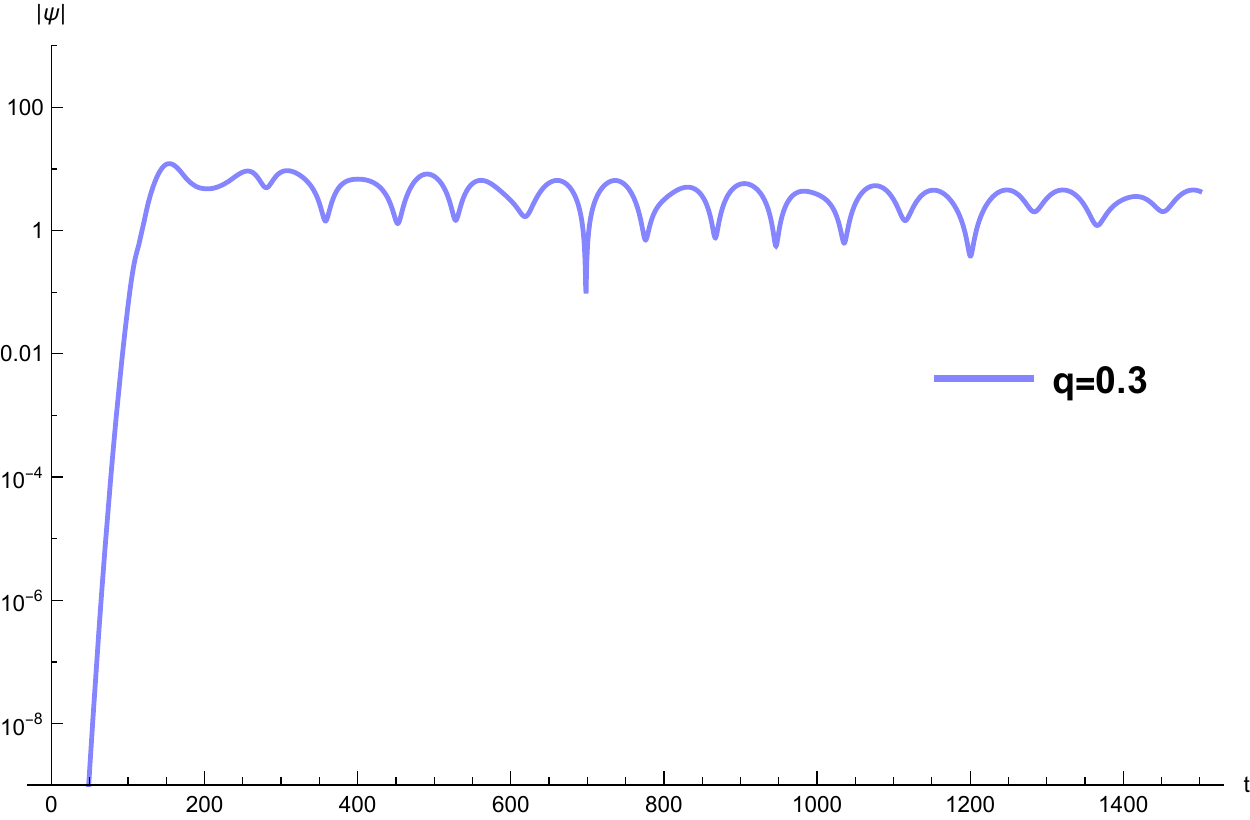}$\quad$\includegraphics[scale=0.6]{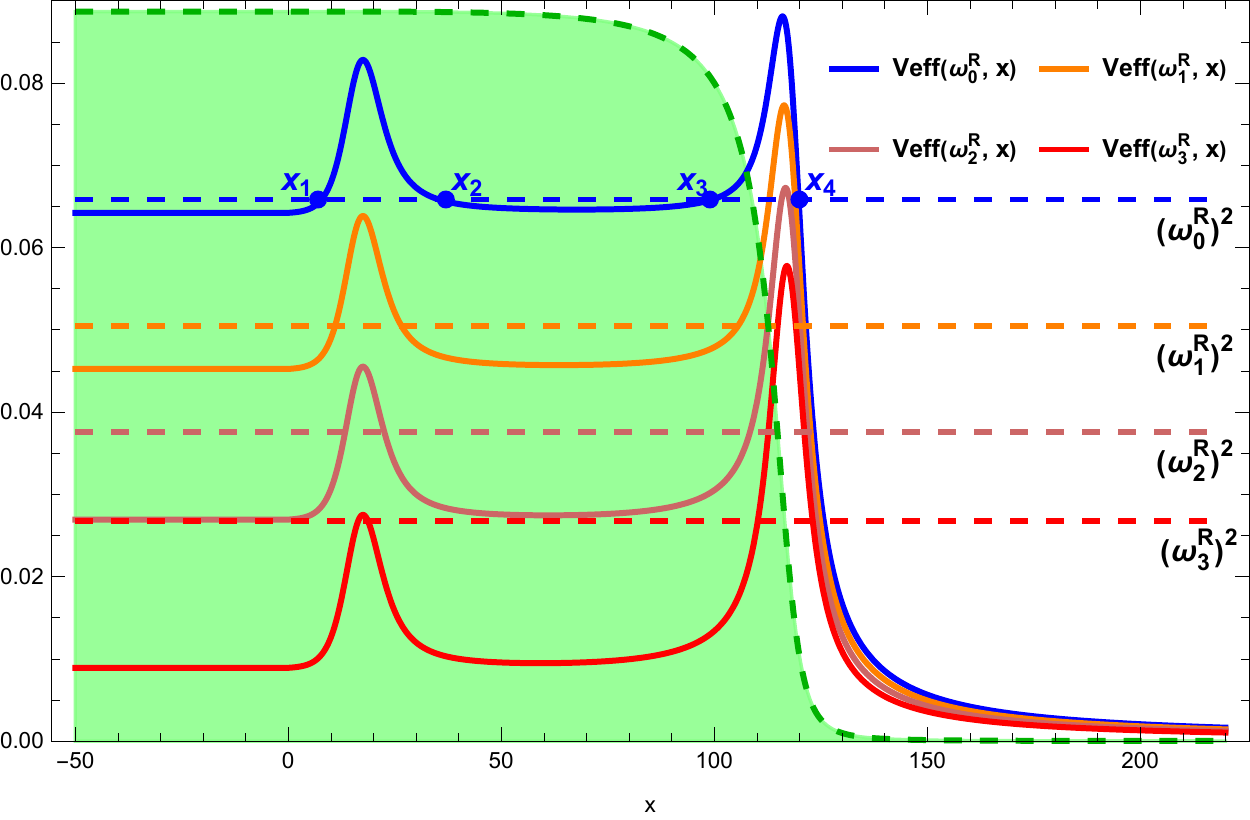}
\par
$\,$
\par
\includegraphics[width=0.9\textwidth]{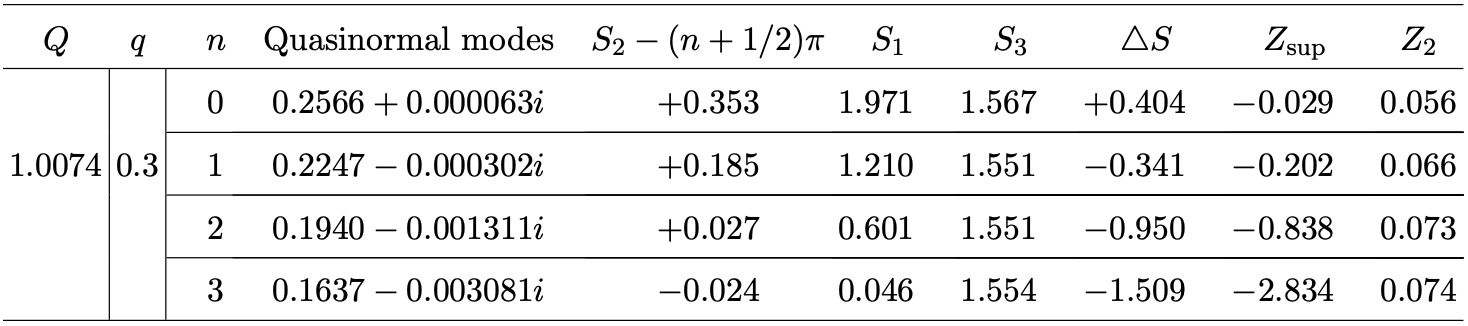}\caption{Superradiance
instability of the $q=0.3$ and $l=0$ scalar field in the hairy black hole with
$\alpha=0.5$ and $Q=1.0074$. \textbf{Upper-Left Panel}: Time evolution of the
scalar field measured at $x=223$. \textbf{Upper-Right Panel}: The effective
potential $V_{\text{eff}}(\omega_{n}^{R},x)$ of the quasinormal modes has two
well-separated peaks. The green region is bounded by $q^{2}\Phi^{2}\left(
x\right)  $. \textbf{Lower Table}: Quasinormal modes extracted from the time
evolution of the scalar field. The $n=0$ mode is a superradiant mode with
$\omega^{I}>0$ while the rest modes are damped ones with $\omega^{I}<0$. As
explained in the main text, $Z_{\text{sup}}+Z_{2}$ and $\Delta S$ can be used
to measure the competition between the superradiant accumulation in the
potential valley and the leakage of modes through the inner potential barrier.
Interestingly, a superradiant/damped mode has positive/negative $Z_{\text{sup}%
}+Z_{2}$ and $\Delta S$. }%
\label{Fig:FigAlpha05}%
\end{figure}

For a large enough black hole charge, there can exist two peaks in the
effective potential of the scalar field. Depending on the black hole and field
parameters, the separation between the two peaks can be considerably larger
than the Compton wavelength of the scalar field, and the potential changes
very slowly with $x$ between the two peaks. In this case, WKB approximate
solutions of the scalar field can be used in the potential well, which
provides insight into superradiance instabilities of hairy black holes.

In FIG. \ref{Fig:FigAlpha05}, we study the scalar field of $q=0.3$ and $l=0$
in the hairy black hole with $\alpha=0.5$ and $Q=1.0074$. The upper-left panel
shows the time evolution of the scalar field measured at $x=$ $223$, and the
extracted quasinormal modes are presented in the lower table. For each
quasinormal mode $\omega_{n}=\omega_{n}^{R}+i\omega_{n}^{I}$, the effective
potential $V_{\text{eff}}\left(  \omega_{n}^{R},x\right)  $ is displayed in
the upper-right panel, which shows that $V_{\text{eff}}\left(  \omega_{n}%
^{R},x\right)  $ indeed has two well-separated potential peaks. Moreover, the
green region in the upper-right panel corresponds to $0<\omega<-q\Phi\left(
x\right)  $, and scalar waves entering or leaving the green region can be
superradiantly amplified.

Intriguingly, the imaginary part of the $n=0$ mode is positive, indicating
that the hairy black hole is unstable against this superradiant mode. To
better understand the superradiant mode, we consider a scalar perturbation
initially perturbed between the two potential peaks. Afterwards, scalar waves
are incident on the two potential barriers, and the transmitted waves carry
away the perturbation to the black hole and spatial infinity. As shown in the
upper-right panel of FIG. \ref{Fig:FigAlpha05}, the transmitted part of the
scalar wave incident on the outer potential barrier leaves the green region,
and hence the reflected wave gets superradiantly amplified. The amplification
can be characterized by the amplification factor $Z_{2}$, which is defined in
eqn. $\left(  \ref{eq:Z2}\right)  $. On the other hand, the leakage of the
scalar wave through the inner potential barrier can suppress the superradiant
amplification. To describe the suppression, one can consider the solution
$\psi_{\text{sup}}\left(  x\right)  $,
\begin{align}
\psi_{\text{sup}}(x) &  \sim e^{-ik_{h}x}\text{ for }x\rightarrow
-\infty\text{,}\nonumber\\
\psi_{\text{sup}}(x) &  \sim\mathcal{I}_{\text{sup}}e^{-ik_{m}x}%
+\mathcal{R}_{\text{sup}}e^{+ik_{m}x}\text{ when }x\text{ is between the two
peaks,}%
\end{align}
where $\mathcal{I}_{\text{sup}}e^{-ik_{m}x}/\mathcal{R}_{\text{sup}}%
e^{+ik_{m}x}$ represents the ingoing/outgoing mode with a negative/positive
group velocity. So one can introduce a suppression factor%
\begin{equation}
Z_{\text{sup}}=1-\frac{\left\vert \mathcal{I}_{\text{sup}}\right\vert ^{2}%
}{\left\vert \mathcal{R}_{\text{sup}}\right\vert ^{2}},
\end{equation}
which is negative if $\left\vert \mathcal{R}_{\text{sup}}\right\vert
^{2}<\left\vert \mathcal{I}_{\text{sup}}\right\vert ^{2}$. Note that more
negative $Z_{\text{sup}}$ is, the more the leakage through the inner potential
barrier suppresses the superradiant amplification. Roughly speaking, the
competition between the superradiant amplification and the suppression
determines whether trapped modes between the potential peaks can accumulate.
In the table of FIG. \ref{Fig:FigAlpha05}, we list $Z_{\text{sup}}$ and
$Z_{2}$ for each mode and find that the superradiant mode with $\omega^{I}>0$
has $Z_{\text{sup}}+Z_{2}>0$ while the damped modes with $\omega^{I}<0$ have
$Z_{\text{sup}}+Z_{2}<0$.

Alternatively for a quasinormal mode $\omega_{n}=\omega_{n}^{R}+i\omega
_{n}^{I}$, we define
\begin{align}
S_{1}  &  =\int_{x_{1}}^{x_{2}}\sqrt{V_{\text{eff}}(\omega_{n}^{R}%
,x)-(\omega_{n}^{R})^{2}}dx,\nonumber\\
S_{2}  &  =\int_{x_{2}}^{x_{3}}\sqrt{(\omega_{n}^{R})^{2}-V_{\text{eff}%
}(\omega_{n}^{R},x)}dx,\\
S_{3}  &  =\int_{x_{3}}^{x_{4}}\sqrt{V_{\text{eff}}(\omega_{n}^{R}%
,x)-(\omega_{n}^{R})^{2}}dx,\nonumber
\end{align}
where $x_{1}$, $x_{2}$, $x_{3}$ and $x_{4}$ are displayed in the upper-right
panel of FIG. \ref{Fig:FigAlpha05}. As expected, a larger $S_{1}$ makes
tunneling through the inner potential barrier more difficult and hence reduces
the leakage of trapped modes. On the other hand, a larger $S_{3}$ makes
tunneling through the outer potential barrier more difficult, meaning that
less scalar waves with positive energy can escape to spatial infinity. So the
superradiance accumulates less trapped modes with negative energy in the
potential valley for a larger $S_{3}$. In the table of FIG.
\ref{Fig:FigAlpha05}, $S_{1}$, $S_{3}$ and $\bigtriangleup S\equiv S_{1}%
-S_{3}$ are presented for each mode. It shows that the sign of $\bigtriangleup
S$ can also be used to reflect which effect wins the competition.
Particularly, a positive $\bigtriangleup S$ implies that the superradiant
amplification plays a more dominant role, therefore signaling a superradiant
mode. Moreover in the eikonal limit, quasinormal modes trapped in the
potential valley were found to satisfy the quantization rule
\cite{Guo:2021enm},
\begin{equation}
S_{2}\approx\left(  n+\frac{1}{2}\right)  \pi,\qquad n=0,1,2\ldots.
\label{eq:WKBBS}%
\end{equation}
Indeed, the table of FIG. \ref{Fig:FigAlpha05} shows that, for each mode, the
numerical values of $S_{2}$ are well approximated by the WKB result $\left(
\ref{eq:WKBBS}\right)  $.

\subsection{Potential with Two Adjacent Peaks}

When the separation between the two potenial peaks are not large enough, one
cannot obtain $Z_{\text{sup}}$ and $Z_{2}$ to discuss the amplification and
suppression effects. Nevertheless, $\bigtriangleup S$ can still be employed to
describe the competition between the superradiant amplification and the
leakage of trapped modes. Roughly speaking, more scalar waves escape to black
holes for a smaller $S_{1}$, and more scalar perturbations are accumulated in
the potential valley for a smaller $S_{3}$. Our results show that, if
$\bigtriangleup S$ is positive, the superradiance effect can compensate the
loss of modes leaking into black holes. So trapped modes can accumulate in the
potential well, and eventually become superradiantly unstable.

\begin{figure}[ptb]
\includegraphics[scale=0.6]{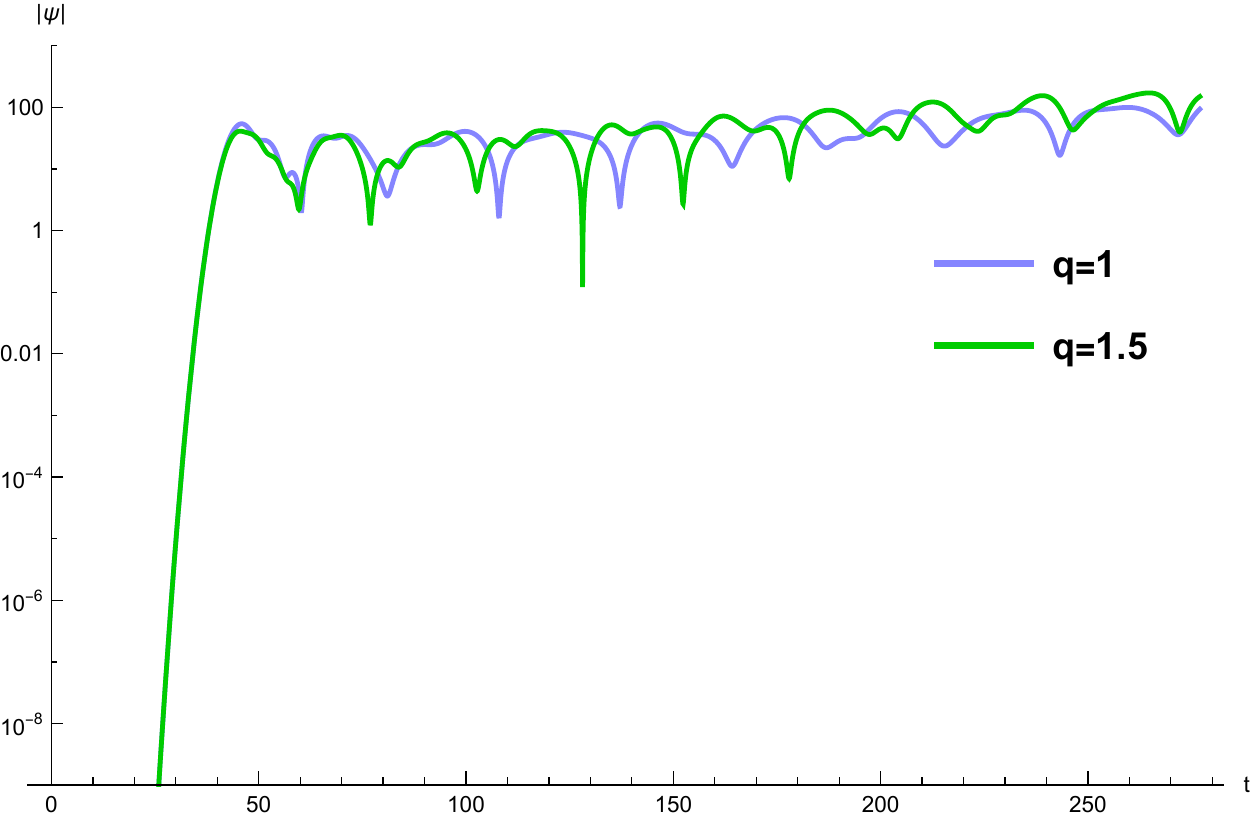}$\quad$%
\includegraphics[scale=0.6]{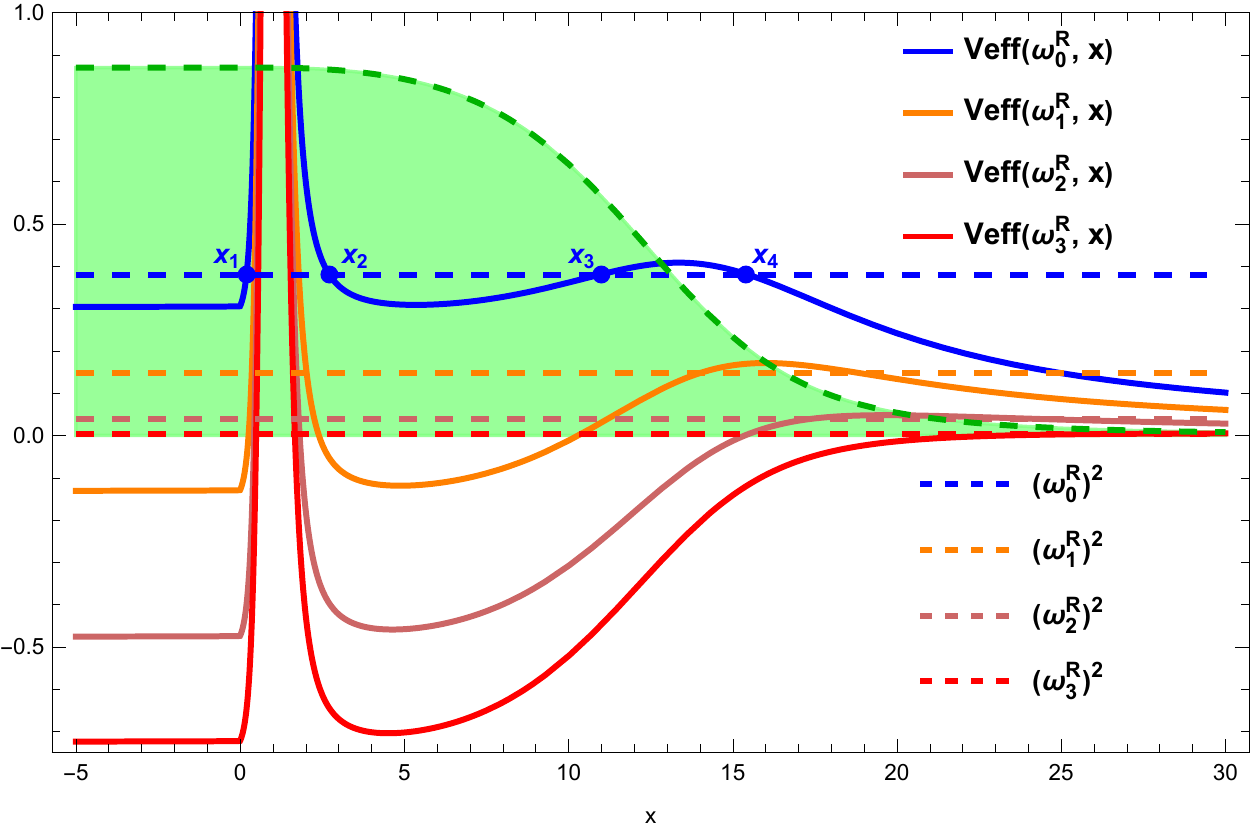}
\par
$\,$
\par
\includegraphics[width=0.7\textwidth]{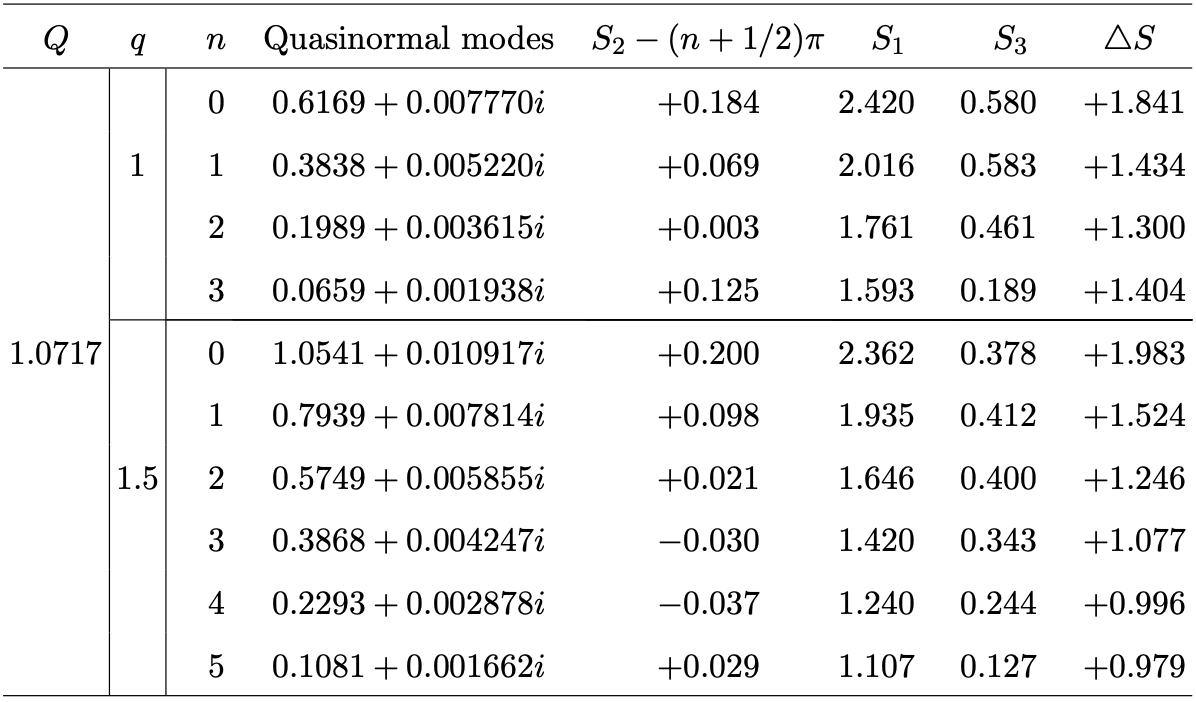}\caption{\textbf{Upper-Left
Panel}: Time evolution of the $l=0$ scalar fields with $q=1$ and $1.5$ in the
hairy black hole with $\alpha=0.88$ and $Q=1.0717$. The observer is located at
$x=56$. The scalar fields grow at late times, and hence the black hole is
unstable against the scalar perturbations. \textbf{Upper-Right Panel}: The
effective potential $V_{\text{eff}}(\omega_{n}^{R},x)$ of the $q=1$ scalar
field has two adjacent potential peaks. \textbf{Lower Table}: All the
extracted quasinormal modes are unstable modes and have a positive $\Delta
S$.}%
\label{Fig:Hairy type}%
\end{figure}

In FIG. \ref{Fig:Hairy type}, we present time evolutions, effective potentials
and quasinormal modes of the monopole scalar fields with $q=1$ and $1.5$ in
the hairy black hole with $\alpha=0.88$ and $Q=1.0717$. Since the $q=1$ and
$1.5$ scalar fields have similar effective potentials, we only display the
$q=1$ case in the upper-right panel. Note that the potential valley is mostly
in the green region. When scalar waves tunnel through the outer potential
barrier and leave the green region, modes trapped in the potential valley will
get superradiantly amplified. The upper-left panel exhibits that the
perturbations grow at late times, indicating an instability of the black hole.
The extracted quasinormal modes are listed in the table of FIG.
\ref{Fig:Hairy type}, and all these modes are found to be superradiant mode
with positive imaginary parts. With a given $n$, the scalar field with $q=1.5$
has a larger imaginary part of the quasinormal mode than that with $q=1$. This
observation is expected since superradiant amplifications become stronger at a
larger $qQ$. Moreover, the superradiant modes all have a positive
$\bigtriangleup S$ and approximately satisfy the quantization rule $\left(
\ref{eq:WKBBS}\right)  $.

\begin{figure}[ptb]
\includegraphics[scale=0.6]{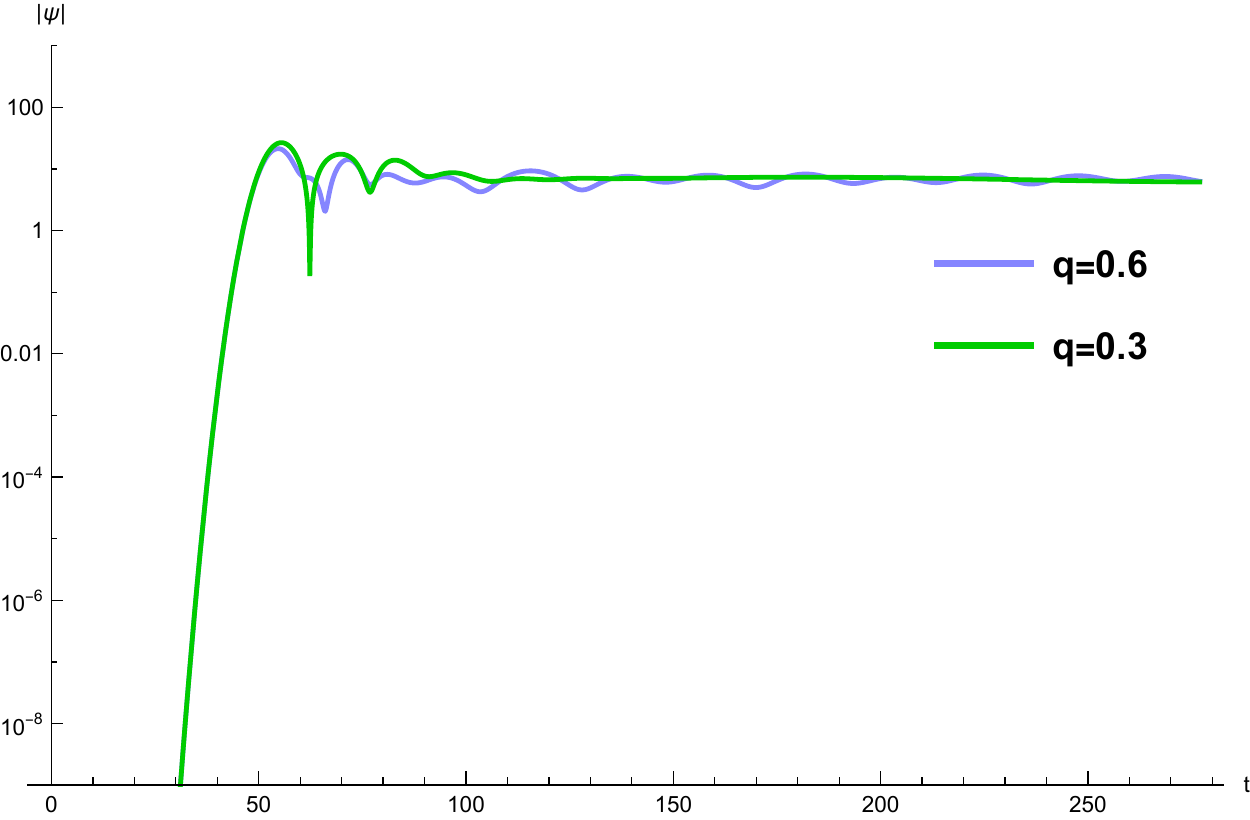}$\quad$%
\includegraphics[scale=0.6]{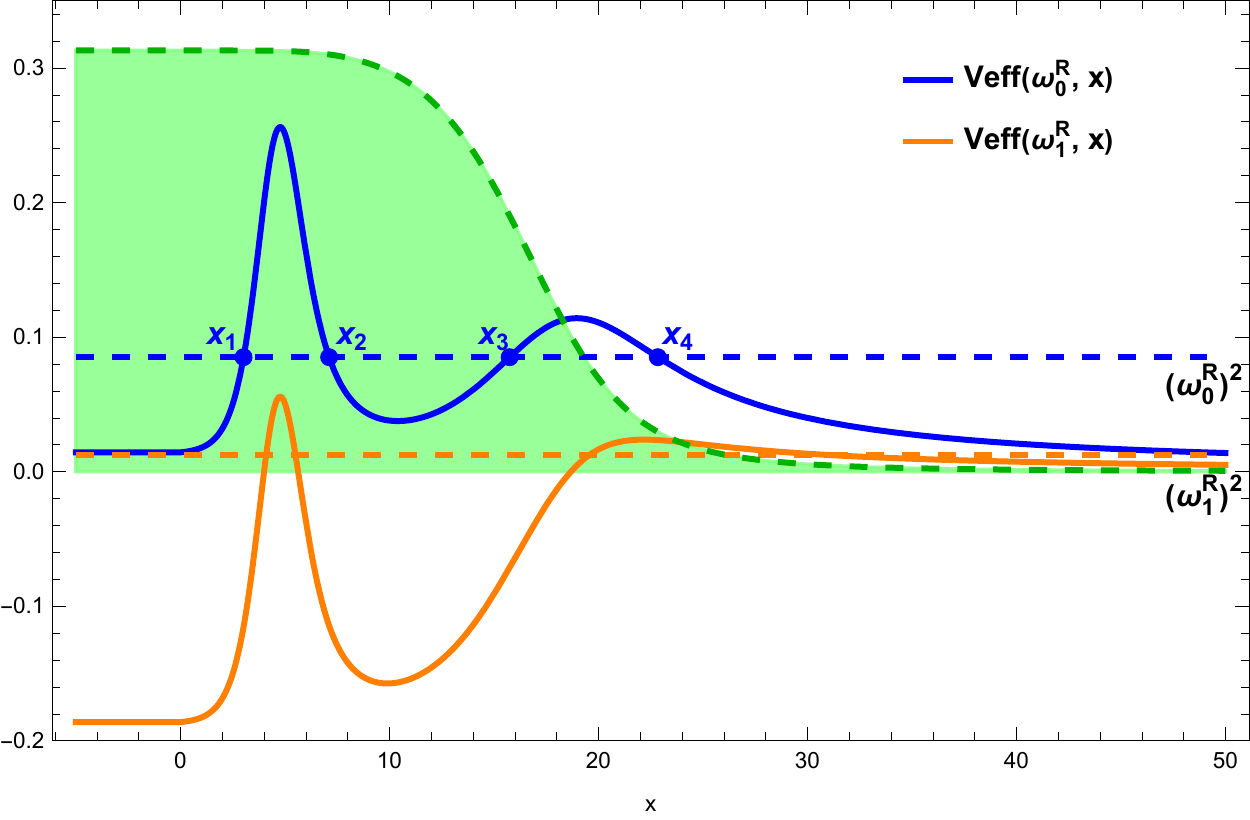}
\par
$\,$
\par
\includegraphics[width=0.7\textwidth]{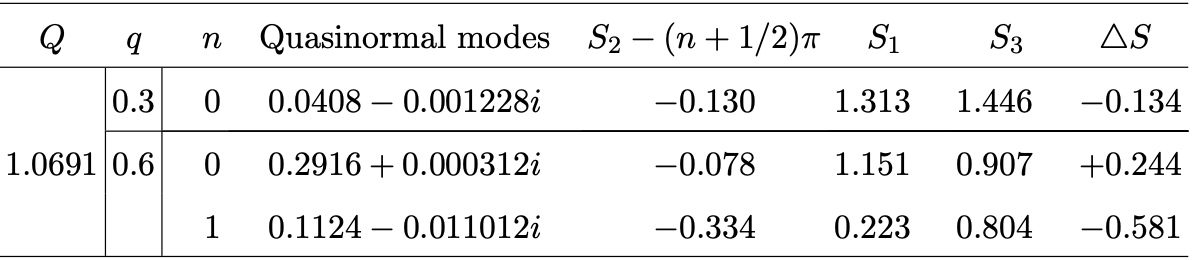}\caption{\textbf{Upper-Left
Panel}: Time evolution of the $l=0$ scalar fields with $q=0.3$ and $0.6$ in
the hairy black hole with $\alpha=0.88$ and $Q=1.0691$. The observer is
located at $x=66$. \textbf{Upper-Right Panel}: The effective potential
$V_{\text{eff}}(\omega_{n}^{R},x)$ of the $q=0.6$ scalar field has a potential
well. \textbf{Lower Table}: The $q=0.6$ scalar field has one unstable
quasinormal mode, which has a positive $\Delta S$.}%
\label{Fig:Hairy-RN R}%
\end{figure}

\begin{figure}[ptb]
\includegraphics[scale=0.6]{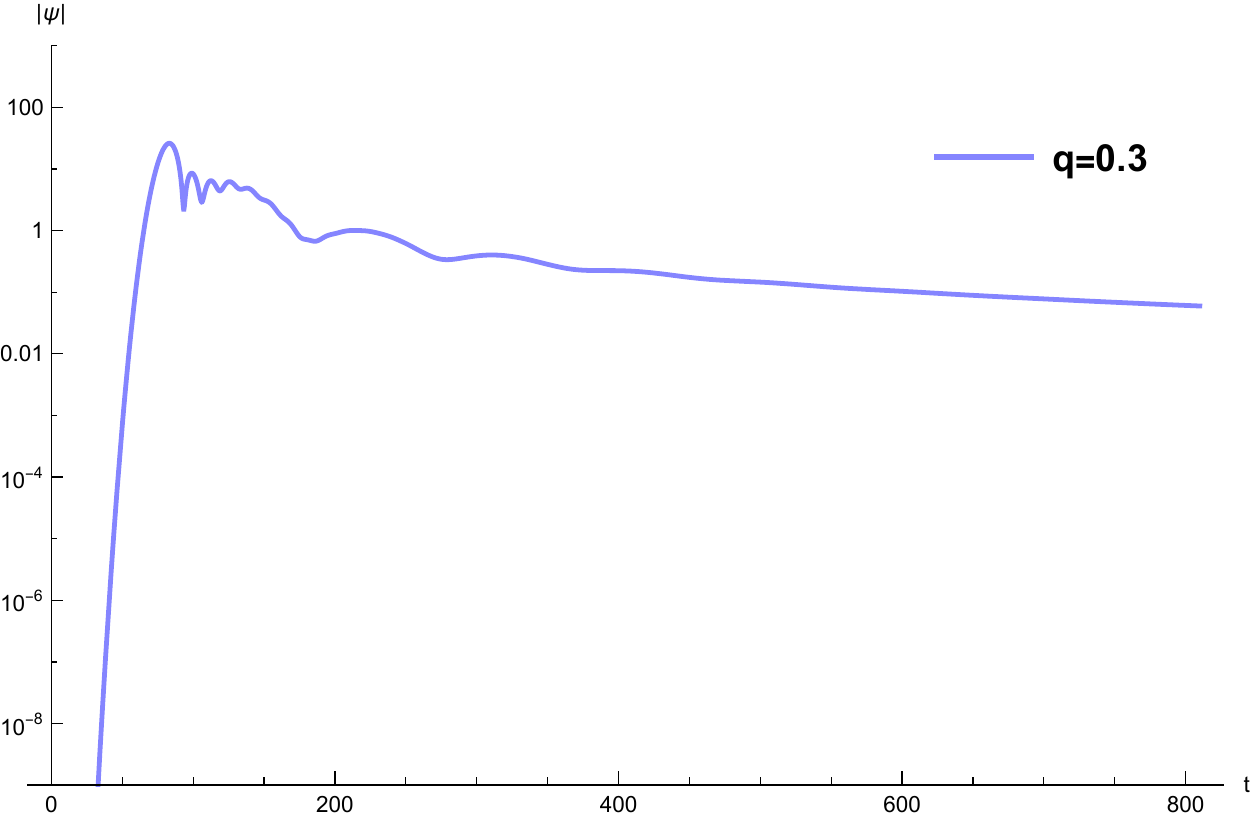}$\quad$%
\includegraphics[scale=0.6]{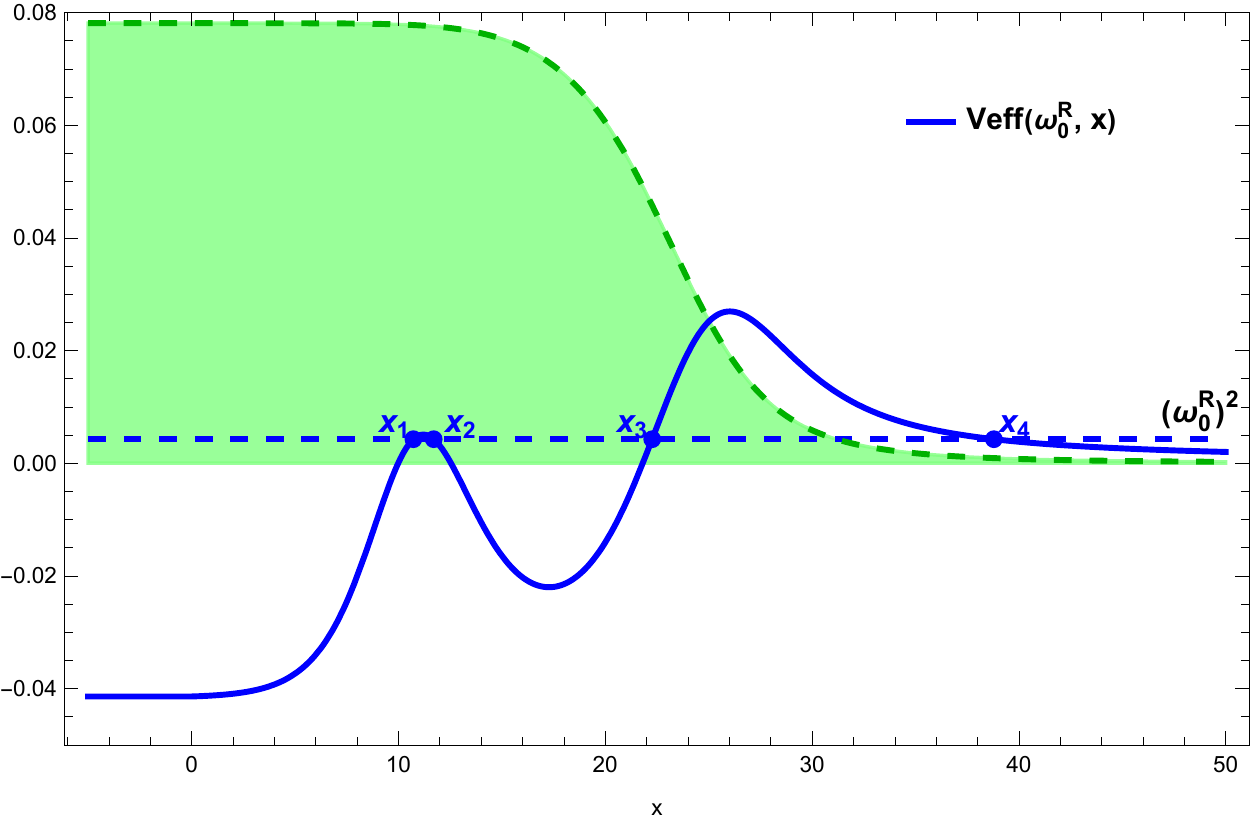}
\par
$\,$
\par
\includegraphics[width=0.7\textwidth]{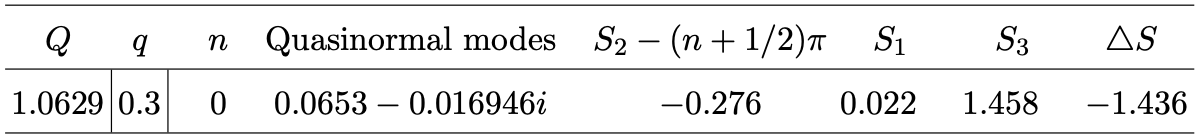}\caption{Time evolution at
$x=105$, effective potential and quasinormal mode of the $l=0$ and $q=0.3$
scalar perturbation in the hairy black hole with $\alpha=0.88$ and $Q=1.0629$.
Due to a high tunneling rate through the inner potential barrier, only one
stable mode is found.}%
\label{Fig:Hairy-RN L}%
\end{figure}

In FIG. \ref{Fig:Hairy-RN R}, we consider the scalar perturbations with the
multipole number $l=0$ and the charge $q=0.3$ and $0.6$ around the hairy black
hole with $\alpha=0.88$ and $Q=1.0691$. Only one unstable mode, i.e.,
$\omega_{0}$ of the $q=0.6$ scalar field, is found, and its effective
potential is displayed as a blue curve in the upper-right panel. In addition,
$\bigtriangleup S$ of the stable and unstable modes is negative and positive,
respectively. FIG. \ref{Fig:Hairy-RN L} discusses the scalar field with $l=0$
and $q=0.3$ in the hairy black hole with $\alpha=0.88$ and $Q=1.0629$. Owing
to a small $S_{1}$, most of trapped modes can tunnel through the inner barrier
and escape to the black hole. Therefore, there is no superradiant mode, and
only a stable mode is obtained. Again, this stable mode has a negative
$\bigtriangleup S$.

\section{Conclusions}

\label{sec:CONCLUSIONS}

In this paper, we investigated superradiance and superradiant instabilities of
charged scalar perturbations in hairy black hole spacetime, where a real
scalar field is minimally coupled to the gravity sector and non-minimally
coupled to the electromagnetic field with an exponential coupling function. It
showed that a charged scalar wave of frequency $\omega$ can be superradiantly
amplified when $\omega<-q\Phi(r_{h})$. Moreover, we found that the energy of
the scalar mode measured by a local observer is negative if $\omega<-q\Phi
(r)$. In the figures of this paper, the green regions represent scalar modes
with $\omega<-q\Phi(r)$, which are negative energy states. When the
transmitted waves enter/leave the green regions, they carry away
negative/positive energies to the black hole/spatial infinity. Hence, the
reflected waves with positive/negative energy are superradiantly amplified.

To trigger a superradiant instability, two ingredients, i.e., the existence of
a trapping potential well and superradiant amplification of trapped modes, are
required. Interestingly, we observed that, in certain parameter regimes, the
effective potential of the scalar field can develop a potential well between
two potential peaks outside the black hole, which can trap some quasinormal
modes. If the trapped modes are in the green regions, they are negative energy
states in the potential well. The transmitted waves tunneling through the
outer potential barrier superradiantly amplify the trapped modes by
transporting positive energy to spatial infinity. On the contrary, the
transmitted waves tunneling through the inner potential barrier carry away
negative energy to black holes and hence tend to suppress the superradiant
amplification. The competition between the two effects would determine whether
the trapped modes can grow and destabilize the black hole spacetime.
Remarkably, we discovered the existence of superradiant quasinormal modes with
a positive imaginary part of the frequency, which renders the black hole
spacetime unstable. In addition, it showed that $\Delta S$ can be used to
describe the competition between the amplification and suppression.
Specifically, our results revealed that unstable and stable modes have
positive and negative $\Delta S$, respectively.

It should be emphasized that the trapping potential well of a massive scalar
field on the Kerr background lies outside the ergoregion, where negative
energy states exist \cite{Arvanitaki:2009fg}. So trapped superradiant modes
are positive energy states in the potential well. Increasing the width or
height of the inner potential barrier leads to the decrease of the transmitted
wave transporting negative energy to the Kerr black hole, which alleviates the
superradiant instability. However in the charged hairy black holes, a smaller
transmission coefficient through the inner potential barrier favors the
accumulation of trapped modes with negative energy, and hence aggravates the
superradiant instability.

Finally, it is desirable to explore the possible end state of the hairy black
holes under charged perturbations. Beyond the linear analysis\ in this paper,
a full nonlinear analysis is necessary to track evolutions of the background
metric, which provides a further understanding on how trapped superradiant
modes affect underlying spacetime. Lately, the \textquotedblleft trapping
instability\textquotedblright\ induced by stable light rings of ultracompact
objects has been confirmed, and two end states of ultracompact bosonic stars
have been identified \cite{Cunha:2022gde,Zhong:2022jke}. The fully nonlinear
numerical evolutions of hairy black holes suffered from superradiant
instabilities may shed light on the fate of stable photon spheres.

\begin{acknowledgments}
We are grateful to Qingyu Gan and Xin Jiang for useful discussions and
valuable comments. This work is supported in part by NSFC (Grant No. 12105191,
12275183, 12275184 and 11875196). Houwen Wu is supported by the International
Visiting Program for Excellent Young Scholars of Sichuan University.
\end{acknowledgments}

\bibliographystyle{unsrturl}
\bibliography{ref}

\begin{thebibliography}{10}

\bibitem{Abbott:2016blz}
B.P. Abbott et~al.
\newblock {Observation of Gravitational Waves from a Binary Black Hole Merger}.
\newblock {\em Phys. Rev. Lett.}, 116(6):061102, 2016.
\newblock \href {http://arxiv.org/abs/1602.03837} {\path{arXiv:1602.03837}},
  \href {http://dx.doi.org/10.1103/PhysRevLett.116.061102}
  {\path{doi:10.1103/PhysRevLett.116.061102}}.

\bibitem{Berti:2007dg}
Emanuele Berti, Vitor Cardoso, Jose~A. Gonzalez, and Ulrich Sperhake.
\newblock {Mining information from binary black hole mergers: A Comparison of
  estimation methods for complex exponentials in noise}.
\newblock {\em Phys. Rev. D}, 75:124017, 2007.
\newblock \href {http://arxiv.org/abs/gr-qc/0701086}
  {\path{arXiv:gr-qc/0701086}}, \href
  {http://dx.doi.org/10.1103/PhysRevD.75.124017}
  {\path{doi:10.1103/PhysRevD.75.124017}}.

\bibitem{Price:2017cjr}
Richard~H. Price and Gaurav Khanna.
\newblock {Gravitational wave sources: reflections and echoes}.
\newblock {\em Class. Quant. Grav.}, 34(22):225005, 2017.
\newblock \href {http://arxiv.org/abs/1702.04833} {\path{arXiv:1702.04833}},
  \href {http://dx.doi.org/10.1088/1361-6382/aa8f29}
  {\path{doi:10.1088/1361-6382/aa8f29}}.

\bibitem{Giesler:2019uxc}
Matthew Giesler, Maximiliano Isi, Mark~A. Scheel, and Saul Teukolsky.
\newblock {Black Hole Ringdown: The Importance of Overtones}.
\newblock {\em Phys. Rev. X}, 9(4):041060, 2019.
\newblock \href {http://arxiv.org/abs/1903.08284} {\path{arXiv:1903.08284}},
  \href {http://dx.doi.org/10.1103/PhysRevX.9.041060}
  {\path{doi:10.1103/PhysRevX.9.041060}}.

\bibitem{Ferrari:1984zz}
Valeria Ferrari and Bahram Mashhoon.
\newblock {New approach to the quasinormal modes of a black hole}.
\newblock {\em Phys. Rev. D}, 30:295--304, 1984.
\newblock \href {http://dx.doi.org/10.1103/PhysRevD.30.295}
  {\path{doi:10.1103/PhysRevD.30.295}}.

\bibitem{Nollert:1999ji}
Hans-Peter Nollert.
\newblock {TOPICAL REVIEW: Quasinormal modes: the characteristic `sound' of
  black holes and neutron stars}.
\newblock {\em Class. Quant. Grav.}, 16:R159--R216, 1999.
\newblock \href {http://dx.doi.org/10.1088/0264-9381/16/12/201}
  {\path{doi:10.1088/0264-9381/16/12/201}}.

\bibitem{Yang:2012he}
Huan Yang, David~A. Nichols, Fan Zhang, Aaron Zimmerman, Zhongyang Zhang, and
  Yanbei Chen.
\newblock {Quasinormal-mode spectrum of Kerr black holes and its geometric
  interpretation}.
\newblock {\em Phys. Rev. D}, 86:104006, 2012.
\newblock \href {http://arxiv.org/abs/1207.4253} {\path{arXiv:1207.4253}},
  \href {http://dx.doi.org/10.1103/PhysRevD.86.104006}
  {\path{doi:10.1103/PhysRevD.86.104006}}.

\bibitem{Konoplya:2017wot}
R.~A. Konoplya and Z.~Stuchl\'\i{}k.
\newblock {Are eikonal quasinormal modes linked to the unstable circular null
  geodesics?}
\newblock {\em Phys. Lett. B}, 771:597--602, 2017.
\newblock \href {http://arxiv.org/abs/1705.05928} {\path{arXiv:1705.05928}},
  \href {http://dx.doi.org/10.1016/j.physletb.2017.06.015}
  {\path{doi:10.1016/j.physletb.2017.06.015}}.

\bibitem{Jusufi:2019ltj}
Kimet Jusufi.
\newblock {Quasinormal Modes of Black Holes Surrounded by Dark Matter and Their
  Connection with the Shadow Radius}.
\newblock {\em Phys. Rev. D}, 101(8):084055, 2020.
\newblock \href {http://arxiv.org/abs/1912.13320} {\path{arXiv:1912.13320}},
  \href {http://dx.doi.org/10.1103/PhysRevD.101.084055}
  {\path{doi:10.1103/PhysRevD.101.084055}}.

\bibitem{Cuadros-Melgar:2020kqn}
B.~Cuadros-Melgar, R.~D.~B. Fontana, and Jeferson de~Oliveira.
\newblock {Analytical correspondence between shadow radius and black hole
  quasinormal frequencies}.
\newblock {\em Phys. Lett. B}, 811:135966, 2020.
\newblock \href {http://arxiv.org/abs/2005.09761} {\path{arXiv:2005.09761}},
  \href {http://dx.doi.org/10.1016/j.physletb.2020.135966}
  {\path{doi:10.1016/j.physletb.2020.135966}}.

\bibitem{Qian:2021aju}
Wei-Liang Qian, Kai Lin, Xiao-Mei Kuang, Bin Wang, and Rui-Hong Yue.
\newblock {Quasinormal modes in two-photon autocorrelation and the
  geometric-optics approximation}.
\newblock 9 2021.
\newblock \href {http://arxiv.org/abs/2109.02844} {\path{arXiv:2109.02844}}.

\bibitem{Cardoso:2008bp}
Vitor Cardoso, Alex~S. Miranda, Emanuele Berti, Helvi Witek, and Vilson~T.
  Zanchin.
\newblock {Geodesic stability, Lyapunov exponents and quasinormal modes}.
\newblock {\em Phys. Rev. D}, 79:064016, 2009.
\newblock \href {http://arxiv.org/abs/0812.1806} {\path{arXiv:0812.1806}},
  \href {http://dx.doi.org/10.1103/PhysRevD.79.064016}
  {\path{doi:10.1103/PhysRevD.79.064016}}.

\bibitem{Cardoso:2017soq}
Vitor Cardoso, Jo\~ao~L. Costa, Kyriakos Destounis, Peter Hintz, and Aron
  Jansen.
\newblock {Quasinormal modes and Strong Cosmic Censorship}.
\newblock {\em Phys. Rev. Lett.}, 120(3):031103, 2018.
\newblock \href {http://arxiv.org/abs/1711.10502} {\path{arXiv:1711.10502}},
  \href {http://dx.doi.org/10.1103/PhysRevLett.120.031103}
  {\path{doi:10.1103/PhysRevLett.120.031103}}.

\bibitem{Gan:2019jac}
Qingyu Gan, Guangzhou Guo, Peng Wang, and Houwen Wu.
\newblock {Strong cosmic censorship for a scalar field in a
  Born-Infeld\textendash{}de Sitter black hole}.
\newblock {\em Phys. Rev. D}, 100(12):124009, 2019.
\newblock \href {http://arxiv.org/abs/1907.04466} {\path{arXiv:1907.04466}},
  \href {http://dx.doi.org/10.1103/PhysRevD.100.124009}
  {\path{doi:10.1103/PhysRevD.100.124009}}.

\bibitem{Gan:2019ibg}
Qingyu Gan, Peng Wang, Houwen Wu, and Haitang Yang.
\newblock {Strong Cosmic Censorship for a Scalar Field in an
  Einstein-Maxwell-Gauss-Bonnet-de Sitter Black Hole}.
\newblock {\em Chin. Phys. C}, 45(2):025103, 2021.
\newblock \href {http://arxiv.org/abs/1911.10996} {\path{arXiv:1911.10996}},
  \href {http://dx.doi.org/10.1088/1674-1137/abccaf}
  {\path{doi:10.1088/1674-1137/abccaf}}.

\bibitem{Berti:2003ud}
E.~Berti and K.~D. Kokkotas.
\newblock {Quasinormal modes of Reissner-Nordstr\"om-anti-de Sitter black
  holes: Scalar, electromagnetic and gravitational perturbations}.
\newblock {\em Phys. Rev. D}, 67:064020, 2003.
\newblock \href {http://arxiv.org/abs/gr-qc/0301052}
  {\path{arXiv:gr-qc/0301052}}, \href
  {http://dx.doi.org/10.1103/PhysRevD.67.064020}
  {\path{doi:10.1103/PhysRevD.67.064020}}.

\bibitem{Berti:2009kk}
Emanuele Berti, Vitor Cardoso, and Andrei~O. Starinets.
\newblock {Quasinormal modes of black holes and black branes}.
\newblock {\em Class. Quant. Grav.}, 26:163001, 2009.
\newblock \href {http://arxiv.org/abs/0905.2975} {\path{arXiv:0905.2975}},
  \href {http://dx.doi.org/10.1088/0264-9381/26/16/163001}
  {\path{doi:10.1088/0264-9381/26/16/163001}}.

\bibitem{Konoplya:2011qq}
R.~A. Konoplya and A.~Zhidenko.
\newblock {Quasinormal modes of black holes: From astrophysics to string
  theory}.
\newblock {\em Rev. Mod. Phys.}, 83:793--836, 2011.
\newblock \href {http://arxiv.org/abs/1102.4014} {\path{arXiv:1102.4014}},
  \href {http://dx.doi.org/10.1103/RevModPhys.83.793}
  {\path{doi:10.1103/RevModPhys.83.793}}.

\bibitem{Cook:2016fge}
Gregory~B. Cook and Maxim Zalutskiy.
\newblock {Purely imaginary quasinormal modes of the Kerr geometry}.
\newblock {\em Class. Quant. Grav.}, 33(24):245008, 2016.
\newblock \href {http://arxiv.org/abs/1603.09710} {\path{arXiv:1603.09710}},
  \href {http://dx.doi.org/10.1088/0264-9381/33/24/245008}
  {\path{doi:10.1088/0264-9381/33/24/245008}}.

\bibitem{Konoplya:2019hlu}
R.~A. Konoplya, A.~Zhidenko, and A.~F. Zinhailo.
\newblock {Higher order WKB formula for quasinormal modes and grey-body
  factors: recipes for quick and accurate calculations}.
\newblock {\em Class. Quant. Grav.}, 36:155002, 2019.
\newblock \href {http://arxiv.org/abs/1904.10333} {\path{arXiv:1904.10333}},
  \href {http://dx.doi.org/10.1088/1361-6382/ab2e25}
  {\path{doi:10.1088/1361-6382/ab2e25}}.

\bibitem{Brito:2015oca}
Richard Brito, Vitor Cardoso, and Paolo Pani.
\newblock {\em {Superradiance}: {New Frontiers in Black Hole Physics}}, volume
  906.
\newblock Springer, 2015.
\newblock \href {http://arxiv.org/abs/1501.06570} {\path{arXiv:1501.06570}},
  \href {http://dx.doi.org/10.1007/978-3-319-19000-6}
  {\path{doi:10.1007/978-3-319-19000-6}}.

\bibitem{Klein:1929zz}
O.~Klein.
\newblock {Die Reflexion von Elektronen an einem Potentialsprung nach der
  relativistischen Dynamik von Dirac}.
\newblock {\em Z. Phys.}, 53:157, 1929.
\newblock \href {http://dx.doi.org/10.1007/BF01339716}
  {\path{doi:10.1007/BF01339716}}.

\bibitem{Cardoso:2012zn}
Vitor Cardoso and Paolo Pani.
\newblock {Tidal acceleration of black holes and superradiance}.
\newblock {\em Class. Quant. Grav.}, 30:045011, 2013.
\newblock \href {http://arxiv.org/abs/1205.3184} {\path{arXiv:1205.3184}},
  \href {http://dx.doi.org/10.1088/0264-9381/30/4/045011}
  {\path{doi:10.1088/0264-9381/30/4/045011}}.

\bibitem{Penrose:1969pc}
R.~Penrose.
\newblock {Gravitational collapse: The role of general relativity}.
\newblock {\em Riv. Nuovo Cim.}, 1:252--276, 1969.
\newblock \href {http://dx.doi.org/10.1023/A:1016578408204}
  {\path{doi:10.1023/A:1016578408204}}.

\bibitem{DiMenza:2014vpa}
Laurent Di~Menza and Jean-Philippe Nicolas.
\newblock {Superradiance on the Reissner\textendash{}Nordstr\o{}m metric}.
\newblock {\em Class. Quant. Grav.}, 32(14):145013, 2015.
\newblock \href {http://arxiv.org/abs/1411.3988} {\path{arXiv:1411.3988}},
  \href {http://dx.doi.org/10.1088/0264-9381/32/14/145013}
  {\path{doi:10.1088/0264-9381/32/14/145013}}.

\bibitem{Zhu:2014sya}
Zhiying Zhu, Shao-Jun Zhang, C.~E. Pellicer, Bin Wang, and Elcio Abdalla.
\newblock {Stability of Reissner-Nordstr\"om black hole in de Sitter background
  under charged scalar perturbation}.
\newblock {\em Phys. Rev. D}, 90(4):044042, 2014.
\newblock [Addendum: Phys.Rev.D 90, 049904 (2014)].
\newblock \href {http://arxiv.org/abs/1405.4931} {\path{arXiv:1405.4931}},
  \href {http://dx.doi.org/10.1103/PhysRevD.90.044042}
  {\path{doi:10.1103/PhysRevD.90.044042}}.

\bibitem{Benone:2015bst}
Carolina~L. Benone and Lu\'\i{}s C.~B. Crispino.
\newblock {Superradiance in static black hole spacetimes}.
\newblock {\em Phys. Rev. D}, 93(2):024028, 2016.
\newblock \href {http://arxiv.org/abs/1511.02634} {\path{arXiv:1511.02634}},
  \href {http://dx.doi.org/10.1103/PhysRevD.93.024028}
  {\path{doi:10.1103/PhysRevD.93.024028}}.

\bibitem{Corelli:2021ikv}
Fabrizio Corelli, Taishi Ikeda, and Paolo Pani.
\newblock {Challenging cosmic censorship in Einstein-Maxwell-scalar theory with
  numerically simulated gedanken experiments}.
\newblock {\em Phys. Rev. D}, 104(8):084069, 2021.
\newblock \href {http://arxiv.org/abs/2108.08328} {\path{arXiv:2108.08328}},
  \href {http://dx.doi.org/10.1103/PhysRevD.104.084069}
  {\path{doi:10.1103/PhysRevD.104.084069}}.

\bibitem{Vicente:2018mxl}
Rodrigo Vicente, Vitor Cardoso, and Jorge~C. Lopes.
\newblock {Penrose process, superradiance, and ergoregion instabilities}.
\newblock {\em Phys. Rev. D}, 97(8):084032, 2018.
\newblock \href {http://arxiv.org/abs/1803.08060} {\path{arXiv:1803.08060}},
  \href {http://dx.doi.org/10.1103/PhysRevD.97.084032}
  {\path{doi:10.1103/PhysRevD.97.084032}}.

\bibitem{Arvanitaki:2009fg}
Asimina Arvanitaki, Savas Dimopoulos, Sergei Dubovsky, Nemanja Kaloper, and
  John March-Russell.
\newblock {String Axiverse}.
\newblock {\em Phys. Rev. D}, 81:123530, 2010.
\newblock \href {http://arxiv.org/abs/0905.4720} {\path{arXiv:0905.4720}},
  \href {http://dx.doi.org/10.1103/PhysRevD.81.123530}
  {\path{doi:10.1103/PhysRevD.81.123530}}.

\bibitem{Hawking:1973uf}
S.~W. Hawking and G.~F.~R. Ellis.
\newblock {\em {The Large Scale Structure of Space-Time}}.
\newblock Cambridge Monographs on Mathematical Physics. Cambridge University
  Press, 2 2011.
\newblock \href {http://dx.doi.org/10.1017/CBO9780511524646}
  {\path{doi:10.1017/CBO9780511524646}}.

\bibitem{Chia:2022udn}
Horng~Sheng Chia, Christoffel Doorman, Alexandra Wernersson, Tanja Hinderer,
  and Samaya Nissanke.
\newblock {Self-Interacting Gravitational Atoms in the Strong-Gravity Regime}.
\newblock 12 2022.
\newblock \href {http://arxiv.org/abs/2212.11948} {\path{arXiv:2212.11948}}.

\bibitem{Berti:2004ju}
Emanuele Berti, Vitor Cardoso, and Jose P.~S. Lemos.
\newblock {Quasinormal modes and classical wave propagation in analogue black
  holes}.
\newblock {\em Phys. Rev. D}, 70:124006, 2004.
\newblock \href {http://arxiv.org/abs/gr-qc/0408099}
  {\path{arXiv:gr-qc/0408099}}, \href
  {http://dx.doi.org/10.1103/PhysRevD.70.124006}
  {\path{doi:10.1103/PhysRevD.70.124006}}.

\bibitem{Cardoso:2004hs}
Vitor Cardoso and Oscar J.~C. Dias.
\newblock {Small Kerr-anti-de Sitter black holes are unstable}.
\newblock {\em Phys. Rev. D}, 70:084011, 2004.
\newblock \href {http://arxiv.org/abs/hep-th/0405006}
  {\path{arXiv:hep-th/0405006}}, \href
  {http://dx.doi.org/10.1103/PhysRevD.70.084011}
  {\path{doi:10.1103/PhysRevD.70.084011}}.

\bibitem{Cardoso:2004nk}
Vitor Cardoso, Oscar J.~C. Dias, Jose P.~S. Lemos, and Shijun Yoshida.
\newblock {The Black hole bomb and superradiant instabilities}.
\newblock {\em Phys. Rev. D}, 70:044039, 2004.
\newblock [Erratum: Phys.Rev.D 70, 049903 (2004)].
\newblock \href {http://arxiv.org/abs/hep-th/0404096}
  {\path{arXiv:hep-th/0404096}}, \href
  {http://dx.doi.org/10.1103/PhysRevD.70.049903}
  {\path{doi:10.1103/PhysRevD.70.049903}}.

\bibitem{Dias:2022str}
Oscar J.~C. Dias, Takaaki Ishii, Keiju Murata, Jorge~E. Santos, and Benson Way.
\newblock {Gregory-Laflamme and Superradiance encounter Black Resonator
  Strings}.
\newblock 12 2022.
\newblock \href {http://arxiv.org/abs/2212.01400} {\path{arXiv:2212.01400}}.

\bibitem{Li:2022hkq}
Zhen Li.
\newblock {Superradiance Instability and Quasinormal Modes of the Gravitational
  Perturbation around Rotating Hairy Black Hole}.
\newblock 12 2022.
\newblock \href {http://arxiv.org/abs/2212.08112} {\path{arXiv:2212.08112}}.

\bibitem{Yang:2022uze}
Hao Yang and Yan-Gang Miao.
\newblock {Superradiance of massive scalar particles around rotating regular
  black holes}.
\newblock 11 2022.
\newblock \href {http://arxiv.org/abs/2211.15130} {\path{arXiv:2211.15130}}.

\bibitem{Hod:2012wmy}
Shahar Hod.
\newblock {Stability of the extremal Reissner-Nordstroem black hole to charged
  scalar perturbations}.
\newblock {\em Phys. Lett. B}, 713:505--508, 2012.
\newblock \href {http://arxiv.org/abs/1304.6474} {\path{arXiv:1304.6474}},
  \href {http://dx.doi.org/10.1016/j.physletb.2012.06.043}
  {\path{doi:10.1016/j.physletb.2012.06.043}}.

\bibitem{Degollado:2013bha}
Juan~Carlos Degollado and Carlos A.~R. Herdeiro.
\newblock {Time evolution of superradiant instabilities for charged black holes
  in a cavity}.
\newblock {\em Phys. Rev. D}, 89(6):063005, 2014.
\newblock \href {http://arxiv.org/abs/1312.4579} {\path{arXiv:1312.4579}},
  \href {http://dx.doi.org/10.1103/PhysRevD.89.063005}
  {\path{doi:10.1103/PhysRevD.89.063005}}.

\bibitem{Herdeiro:2013pia}
Carlos A.~R. Herdeiro, Juan~Carlos Degollado, and Helgi~Freyr R\'unarsson.
\newblock {Rapid growth of superradiant instabilities for charged black holes
  in a cavity}.
\newblock {\em Phys. Rev. D}, 88:063003, 2013.
\newblock \href {http://arxiv.org/abs/1305.5513} {\path{arXiv:1305.5513}},
  \href {http://dx.doi.org/10.1103/PhysRevD.88.063003}
  {\path{doi:10.1103/PhysRevD.88.063003}}.

\bibitem{Dolan:2015dha}
Sam~R Dolan, Supakchai Ponglertsakul, and Elizabeth Winstanley.
\newblock {Stability of black holes in Einstein-charged scalar field theory in
  a cavity}.
\newblock {\em Phys. Rev. D}, 92(12):124047, 2015.
\newblock \href {http://arxiv.org/abs/1507.02156} {\path{arXiv:1507.02156}},
  \href {http://dx.doi.org/10.1103/PhysRevD.92.124047}
  {\path{doi:10.1103/PhysRevD.92.124047}}.

\bibitem{Sanchis-Gual:2016tcm}
Nicolas Sanchis-Gual, Juan~Carlos Degollado, Carlos Herdeiro, José~A. Font,
  and Pedro~J. Montero.
\newblock {Dynamical formation of a Reissner-Nordström black hole with scalar
  hair in a cavity}.
\newblock {\em Phys. Rev. D}, 94(4):044061, 2016.
\newblock \href {http://arxiv.org/abs/1607.06304} {\path{arXiv:1607.06304}},
  \href {http://dx.doi.org/10.1103/PhysRevD.94.044061}
  {\path{doi:10.1103/PhysRevD.94.044061}}.

\bibitem{Kolyvaris:2018zxl}
Theodoros Kolyvaris, Marina Koukouvaou, Antri Machattou, and Eleftherios
  Papantonopoulos.
\newblock {Superradiant instabilities in scalar-tensor Horndeski theory}.
\newblock {\em Phys. Rev. D}, 98(2):024045, 2018.
\newblock \href {http://arxiv.org/abs/1806.11110} {\path{arXiv:1806.11110}},
  \href {http://dx.doi.org/10.1103/PhysRevD.98.024045}
  {\path{doi:10.1103/PhysRevD.98.024045}}.

\bibitem{Huang:2015cha}
Yang Huang and Dao-Jun Liu.
\newblock {Charged scalar perturbations around a regular magnetic black hole}.
\newblock {\em Phys. Rev. D}, 93(10):104011, 2016.
\newblock \href {http://arxiv.org/abs/1509.09017} {\path{arXiv:1509.09017}},
  \href {http://dx.doi.org/10.1103/PhysRevD.93.104011}
  {\path{doi:10.1103/PhysRevD.93.104011}}.

\bibitem{Herdeiro:2018wub}
Carlos~A.R. Herdeiro, Eugen Radu, Nicolas Sanchis-Gual, and Jos\'e~A. Font.
\newblock {Spontaneous Scalarization of Charged Black Holes}.
\newblock {\em Phys. Rev. Lett.}, 121(10):101102, 2018.
\newblock \href {http://arxiv.org/abs/1806.05190} {\path{arXiv:1806.05190}},
  \href {http://dx.doi.org/10.1103/PhysRevLett.121.101102}
  {\path{doi:10.1103/PhysRevLett.121.101102}}.

\bibitem{Fernandes:2019rez}
Pedro G.~S. Fernandes, Carlos A.~R. Herdeiro, Alexandre~M. Pombo, Eugen Radu,
  and Nicolas Sanchis-Gual.
\newblock {Spontaneous Scalarisation of Charged Black Holes: Coupling
  Dependence and Dynamical Features}.
\newblock {\em Class. Quant. Grav.}, 36(13):134002, 2019.
\newblock [Erratum: Class.Quant.Grav. 37, 049501 (2020)].
\newblock \href {http://arxiv.org/abs/1902.05079} {\path{arXiv:1902.05079}},
  \href {http://dx.doi.org/10.1088/1361-6382/ab23a1}
  {\path{doi:10.1088/1361-6382/ab23a1}}.

\bibitem{Fernandes:2019kmh}
Pedro~G.S. Fernandes, Carlos~A.R. Herdeiro, Alexandre~M. Pombo, Eugen Radu, and
  Nicolas Sanchis-Gual.
\newblock {Charged black holes with axionic-type couplings: Classes of
  solutions and dynamical scalarization}.
\newblock {\em Phys. Rev. D}, 100(8):084045, 2019.
\newblock \href {http://arxiv.org/abs/1908.00037} {\path{arXiv:1908.00037}},
  \href {http://dx.doi.org/10.1103/PhysRevD.100.084045}
  {\path{doi:10.1103/PhysRevD.100.084045}}.

\bibitem{Blazquez-Salcedo:2020nhs}
Jose~Luis Bl\'azquez-Salcedo, Carlos~A.R. Herdeiro, Jutta Kunz, Alexandre~M.
  Pombo, and Eugen Radu.
\newblock {Einstein-Maxwell-scalar black holes: the hot, the cold and the
  bald}.
\newblock {\em Phys. Lett. B}, 806:135493, 2020.
\newblock \href {http://arxiv.org/abs/2002.00963} {\path{arXiv:2002.00963}},
  \href {http://dx.doi.org/10.1016/j.physletb.2020.135493}
  {\path{doi:10.1016/j.physletb.2020.135493}}.

\bibitem{Zou:2019bpt}
De-Cheng Zou and Yun~Soo Myung.
\newblock {Scalarized charged black holes with scalar mass term}.
\newblock {\em Phys. Rev. D}, 100(12):124055, 2019.
\newblock \href {http://arxiv.org/abs/1909.11859} {\path{arXiv:1909.11859}},
  \href {http://dx.doi.org/10.1103/PhysRevD.100.124055}
  {\path{doi:10.1103/PhysRevD.100.124055}}.

\bibitem{Fernandes:2020gay}
Pedro~G.S. Fernandes.
\newblock {Einstein-Maxwell-scalar black holes with massive and
  self-interacting scalar hair}.
\newblock {\em Phys. Dark Univ.}, 30:100716, 2020.
\newblock \href {http://arxiv.org/abs/2003.01045} {\path{arXiv:2003.01045}},
  \href {http://dx.doi.org/10.1016/j.dark.2020.100716}
  {\path{doi:10.1016/j.dark.2020.100716}}.

\bibitem{Peng:2019cmm}
Yan Peng.
\newblock {Scalarization of horizonless reflecting stars: neutral scalar fields
  non-minimally coupled to Maxwell fields}.
\newblock {\em Phys. Lett. B}, 804:135372, 2020.
\newblock \href {http://arxiv.org/abs/1912.11989} {\path{arXiv:1912.11989}},
  \href {http://dx.doi.org/10.1016/j.physletb.2020.135372}
  {\path{doi:10.1016/j.physletb.2020.135372}}.

\bibitem{Myung:2018vug}
Yun~Soo Myung and De-Cheng Zou.
\newblock {Instability of Reissner\textendash{}Nordstr\"om black hole in
  Einstein-Maxwell-scalar theory}.
\newblock {\em Eur. Phys. J. C}, 79(3):273, 2019.
\newblock \href {http://arxiv.org/abs/1808.02609} {\path{arXiv:1808.02609}},
  \href {http://dx.doi.org/10.1140/epjc/s10052-019-6792-6}
  {\path{doi:10.1140/epjc/s10052-019-6792-6}}.

\bibitem{Myung:2019oua}
Yun~Soo Myung and De-Cheng Zou.
\newblock {Stability of scalarized charged black holes in the
  Einstein\textendash{}Maxwell\textendash{}Scalar theory}.
\newblock {\em Eur. Phys. J. C}, 79(8):641, 2019.
\newblock \href {http://arxiv.org/abs/1904.09864} {\path{arXiv:1904.09864}},
  \href {http://dx.doi.org/10.1140/epjc/s10052-019-7176-7}
  {\path{doi:10.1140/epjc/s10052-019-7176-7}}.

\bibitem{Zou:2020zxq}
De-Cheng Zou and Yun~Soo Myung.
\newblock {Radial perturbations of the scalarized black holes in
  Einstein-Maxwell-conformally coupled scalar theory}.
\newblock {\em Phys. Rev. D}, 102(6):064011, 2020.
\newblock \href {http://arxiv.org/abs/2005.06677} {\path{arXiv:2005.06677}},
  \href {http://dx.doi.org/10.1103/PhysRevD.102.064011}
  {\path{doi:10.1103/PhysRevD.102.064011}}.

\bibitem{Myung:2020etf}
Yun~Soo Myung and De-Cheng Zou.
\newblock {Onset of rotating scalarized black holes in
  Einstein-Chern-Simons-Scalar theory}.
\newblock {\em Phys. Lett. B}, 814:136081, 2021.
\newblock \href {http://arxiv.org/abs/2012.02375} {\path{arXiv:2012.02375}},
  \href {http://dx.doi.org/10.1016/j.physletb.2021.136081}
  {\path{doi:10.1016/j.physletb.2021.136081}}.

\bibitem{Mai:2020sac}
Zhan-Feng Mai and Run-Qiu Yang.
\newblock {Stability analysis of a charged black hole with a nonlinear complex
  scalar field}.
\newblock {\em Phys. Rev. D}, 104(4):044008, 2021.
\newblock \href {http://arxiv.org/abs/2101.00026} {\path{arXiv:2101.00026}},
  \href {http://dx.doi.org/10.1103/PhysRevD.104.044008}
  {\path{doi:10.1103/PhysRevD.104.044008}}.

\bibitem{Astefanesei:2020qxk}
Dumitru Astefanesei, Carlos Herdeiro, Jo\~ao Oliveira, and Eugen Radu.
\newblock {Higher dimensional black hole scalarization}.
\newblock {\em JHEP}, 09:186, 2020.
\newblock \href {http://arxiv.org/abs/2007.04153} {\path{arXiv:2007.04153}},
  \href {http://dx.doi.org/10.1007/JHEP09(2020)186}
  {\path{doi:10.1007/JHEP09(2020)186}}.

\bibitem{Myung:2018jvi}
Yun~Soo Myung and De-Cheng Zou.
\newblock {Quasinormal modes of scalarized black holes in the
  Einstein\textendash{}Maxwell\textendash{}Scalar theory}.
\newblock {\em Phys. Lett. B}, 790:400--407, 2019.
\newblock \href {http://arxiv.org/abs/1812.03604} {\path{arXiv:1812.03604}},
  \href {http://dx.doi.org/10.1016/j.physletb.2019.01.046}
  {\path{doi:10.1016/j.physletb.2019.01.046}}.

\bibitem{Blazquez-Salcedo:2020jee}
Jose~Luis Bl\'azquez-Salcedo, Carlos~A.R. Herdeiro, Sarah Kahlen, Jutta Kunz,
  Alexandre~M. Pombo, and Eugen Radu.
\newblock {Quasinormal modes of hot, cold and bald Einstein-Maxwell-scalar
  black holes}.
\newblock 8 2020.
\newblock \href {http://arxiv.org/abs/2008.11744} {\path{arXiv:2008.11744}}.

\bibitem{Myung:2020dqt}
Yun~Soo Myung and De-Cheng Zou.
\newblock {Scalarized charged black holes in the Einstein-Maxwell-Scalar theory
  with two U(1) fields}.
\newblock {\em Phys. Lett. B}, 811:135905, 2020.
\newblock \href {http://arxiv.org/abs/2009.05193} {\path{arXiv:2009.05193}},
  \href {http://dx.doi.org/10.1016/j.physletb.2020.135905}
  {\path{doi:10.1016/j.physletb.2020.135905}}.

\bibitem{Myung:2020ctt}
Yun~Soo Myung and De-Cheng Zou.
\newblock {Scalarized black holes in the Einstein-Maxwell-scalar theory with a
  quasitopological term}.
\newblock {\em Phys. Rev. D}, 103(2):024010, 2021.
\newblock \href {http://arxiv.org/abs/2011.09665} {\path{arXiv:2011.09665}},
  \href {http://dx.doi.org/10.1103/PhysRevD.103.024010}
  {\path{doi:10.1103/PhysRevD.103.024010}}.

\bibitem{Guo:2020zqm}
Hong Guo, Xiao-Mei Kuang, Eleftherios Papantonopoulos, and Bin Wang.
\newblock {Topology and spacetime structure influences on black hole
  scalarization}.
\newblock 12 2020.
\newblock \href {http://arxiv.org/abs/2012.11844} {\path{arXiv:2012.11844}}.

\bibitem{Brihaye:2019dck}
Yves Brihaye, Betti Hartmann, Nath\'alia~Pio Aprile, and Jon Urrestilla.
\newblock {Scalarization of asymptotically anti\textendash{}de Sitter black
  holes with applications to holographic phase transitions}.
\newblock {\em Phys. Rev. D}, 101(12):124016, 2020.
\newblock \href {http://arxiv.org/abs/1911.01950} {\path{arXiv:1911.01950}},
  \href {http://dx.doi.org/10.1103/PhysRevD.101.124016}
  {\path{doi:10.1103/PhysRevD.101.124016}}.

\bibitem{Brihaye:2019gla}
Yves Brihaye, Carlos Herdeiro, and Eugen Radu.
\newblock {Black Hole Spontaneous Scalarisation with a Positive Cosmological
  Constant}.
\newblock {\em Phys. Lett. B}, 802:135269, 2020.
\newblock \href {http://arxiv.org/abs/1910.05286} {\path{arXiv:1910.05286}},
  \href {http://dx.doi.org/10.1016/j.physletb.2020.135269}
  {\path{doi:10.1016/j.physletb.2020.135269}}.

\bibitem{Zhang:2021etr}
Cheng-Yong Zhang, Peng Liu, Yunqi Liu, Chao Niu, and Bin Wang.
\newblock {Dynamical charged black hole spontaneous scalarization in Anti-de
  Sitter spacetimes}.
\newblock 3 2021.
\newblock \href {http://arxiv.org/abs/2103.13599} {\path{arXiv:2103.13599}}.

\bibitem{Guo:2021zed}
Guangzhou Guo, Peng Wang, Houwen Wu, and Haitang Yang.
\newblock {Scalarized Einstein\textendash{}Maxwell-scalar black holes in
  anti-de Sitter spacetime}.
\newblock {\em Eur. Phys. J. C}, 81(10):864, 2021.
\newblock \href {http://arxiv.org/abs/2102.04015} {\path{arXiv:2102.04015}},
  \href {http://dx.doi.org/10.1140/epjc/s10052-021-09614-7}
  {\path{doi:10.1140/epjc/s10052-021-09614-7}}.

\bibitem{Gan:2021pwu}
Qingyu Gan, Peng Wang, Houwen Wu, and Haitang Yang.
\newblock {Photon spheres and spherical accretion image of a hairy black hole}.
\newblock {\em Phys. Rev. D}, 104(2):024003, 2021.
\newblock \href {http://arxiv.org/abs/2104.08703} {\path{arXiv:2104.08703}},
  \href {http://dx.doi.org/10.1103/PhysRevD.104.024003}
  {\path{doi:10.1103/PhysRevD.104.024003}}.

\bibitem{Gan:2021xdl}
Qingyu Gan, Peng Wang, Houwen Wu, and Haitang Yang.
\newblock {Photon ring and observational appearance of a hairy black hole}.
\newblock {\em Phys. Rev. D}, 104(4):044049, 2021.
\newblock \href {http://arxiv.org/abs/2105.11770} {\path{arXiv:2105.11770}},
  \href {http://dx.doi.org/10.1103/PhysRevD.104.044049}
  {\path{doi:10.1103/PhysRevD.104.044049}}.

\bibitem{Guo:2022muy}
Guangzhou Guo, Xin Jiang, Peng Wang, and Houwen Wu.
\newblock {Gravitational Lensing by Black Holes with Multiple Photon Spheres}.
\newblock 4 2022.
\newblock \href {http://arxiv.org/abs/2204.13948} {\path{arXiv:2204.13948}}.

\bibitem{Chen:2022qrw}
Yiqian Chen, Guangzhou Guo, Peng Wang, Houwen Wu, and Haitang Yang.
\newblock {Appearance of an infalling star in black holes with multiple photon
  spheres}.
\newblock {\em Sci. China Phys. Mech. Astron.}, 65(12):120412, 2022.
\newblock \href {http://arxiv.org/abs/2206.13705} {\path{arXiv:2206.13705}},
  \href {http://dx.doi.org/10.1007/s11433-022-1986-x}
  {\path{doi:10.1007/s11433-022-1986-x}}.

\bibitem{Liu:2019rib}
Hai-Shan Liu, Zhan-Feng Mai, Yue-Zhou Li, and H.~L\"u.
\newblock {Quasi-topological Electromagnetism: Dark Energy, Dyonic Black Holes,
  Stable Photon Spheres and Hidden Electromagnetic Duality}.
\newblock {\em Sci. China Phys. Mech. Astron.}, 63:240411, 2020.
\newblock \href {http://arxiv.org/abs/1907.10876} {\path{arXiv:1907.10876}},
  \href {http://dx.doi.org/10.1007/s11433-019-1446-1}
  {\path{doi:10.1007/s11433-019-1446-1}}.

\bibitem{deRham:2010kj}
Claudia de~Rham, Gregory Gabadadze, and Andrew~J. Tolley.
\newblock {Resummation of Massive Gravity}.
\newblock {\em Phys. Rev. Lett.}, 106:231101, 2011.
\newblock \href {http://arxiv.org/abs/1011.1232} {\path{arXiv:1011.1232}},
  \href {http://dx.doi.org/10.1103/PhysRevLett.106.231101}
  {\path{doi:10.1103/PhysRevLett.106.231101}}.

\bibitem{Dong:2020odp}
Ruifeng Dong and Dejan Stojkovic.
\newblock {Gravitational wave echoes from black holes in massive gravity}.
\newblock {\em Phys. Rev. D}, 103(2):024058, 2021.
\newblock \href {http://arxiv.org/abs/2011.04032} {\path{arXiv:2011.04032}},
  \href {http://dx.doi.org/10.1103/PhysRevD.103.024058}
  {\path{doi:10.1103/PhysRevD.103.024058}}.

\bibitem{Guo:2022ghl}
Guangzhou Guo, Yuhang Lu, Peng Wang, Houwen Wu, and Haitang Yang.
\newblock {Black Holes with Multiple Photon Spheres}.
\newblock 12 2022.
\newblock \href {http://arxiv.org/abs/2212.12901} {\path{arXiv:2212.12901}}.

\bibitem{Guo:2021enm}
Guangzhou Guo, Peng Wang, Houwen Wu, and Haitang Yang.
\newblock {Quasinormal modes of black holes with multiple photon spheres}.
\newblock {\em JHEP}, 06:060, 2022.
\newblock \href {http://arxiv.org/abs/2112.14133} {\path{arXiv:2112.14133}},
  \href {http://dx.doi.org/10.1007/JHEP06(2022)060}
  {\path{doi:10.1007/JHEP06(2022)060}}.

\bibitem{Guo:2022umh}
Guangzhou Guo, Peng Wang, Houwen Wu, and Haitang Yang.
\newblock {Echoes from hairy black holes}.
\newblock {\em JHEP}, 06:073, 2022.
\newblock \href {http://arxiv.org/abs/2204.00982} {\path{arXiv:2204.00982}},
  \href {http://dx.doi.org/10.1007/JHEP06(2022)073}
  {\path{doi:10.1007/JHEP06(2022)073}}.

\bibitem{Konoplya:2013rxa}
R.~A. Konoplya and A.~Zhidenko.
\newblock {Massive charged scalar field in the Kerr-Newman background I:
  quasinormal modes, late-time tails and stability}.
\newblock {\em Phys. Rev. D}, 88:024054, 2013.
\newblock \href {http://arxiv.org/abs/1307.1812} {\path{arXiv:1307.1812}},
  \href {http://dx.doi.org/10.1103/PhysRevD.88.024054}
  {\path{doi:10.1103/PhysRevD.88.024054}}.

\bibitem{Cunha:2022gde}
Pedro V.~P. Cunha, Carlos Herdeiro, Eugen Radu, and Nicolas Sanchis-Gual.
\newblock {The fate of the light-ring instability}.
\newblock 7 2022.
\newblock \href {http://arxiv.org/abs/2207.13713} {\path{arXiv:2207.13713}}.

\bibitem{Zhong:2022jke}
Zhen Zhong, Vitor Cardoso, and Elisa Maggio.
\newblock {On the instability of ultracompact horizonless spacetimes}.
\newblock 11 2022.
\newblock \href {http://arxiv.org/abs/2211.16526} {\path{arXiv:2211.16526}}.

\end{thebibliography}

\end{document}